\documentclass[sigconf]{acmart}
\renewcommand\footnotetextcopyrightpermission[1]{}
\AtBeginDocument{%
  }
\acmConference[Accepted at AsiaCCS]{ACM ASIA Conference on Computer and Communications Security}{2025}{Vietnam}
\usepackage{subcaption}
\usepackage{caption}
\usepackage{float}
\usepackage{titlesec}
\usepackage{multirow}
\usepackage{dblfloatfix}
\usepackage{graphicx}
\usepackage{xspace}

\begin{document}

\title{Nosy Layers, Noisy Fixes: Tackling DRAs in Federated Learning Systems using Explainable AI}

\author{Meghali Nandi}
\affiliation{%
  \institution{University of New South Wales }
  \city{Sydney}
  \country{Australia}
}
\affiliation{%
  \institution{CSIRO's Data61}
  \city{}
  \country{}
}
\email{m.nandi@unsw.edu.au}

\author{Arash Shaghaghi}
\affiliation{%
  \institution{University of New South Wales}
  \city{Sydney}
  \country{Australia}
}
\email{a.shaghaghi@unsw.edu.au}

\author{Nazatul Haque Sultan}
\affiliation{%
  \institution{CSIRO's Data61}
  \city{}
  \country{}
}
\email{nazatul.sultan@data61.csiro.au}

\author{Gustavo Batista}
\affiliation{%
  \institution{University of New South Wales}
  \city{Sydney}
  \country{Australia}
}
\email{g.batista@unsw.edu.au}

\author{Raymond K. Zhao}
\affiliation{%
  \institution{CSIRO's Data61}
  \city{}
  \country{}
}
\email{raymond.zhao@data61.csiro.au}

\author{Sanjay Jha}
\affiliation{%
  \institution{University of New South Wales}
  \city{Sydney}
  \country{Australia}
}
\email{sanjay.jha@unsw.edu.au}

\renewcommand{\shortauthors}{Meghali et al.}
\newcommand{\DRArmor}{{DRArmor}\xspace}

\begin{abstract}
  Federated Learning (FL) has emerged as a powerful paradigm for collaborative model training while keeping client data decentralized and private. However, it is vulnerable to Data Reconstruction Attacks (DRA) such as ``LoKI'' and ``Robbing the Fed'', where malicious models sent from the server to the client can reconstruct sensitive user data. To counter this, we introduce \DRArmor, a novel defense mechanism that integrates Explainable AI with targeted detection and mitigation strategies for DRA. Unlike existing defenses that focus on the entire model, \DRArmor identifies and addresses the root cause (i.e., malicious layers within the model that send gradients with malicious intent) by analyzing their contribution to the output and detecting inconsistencies in gradient values. Once these malicious layers are identified, \DRArmor applies defense techniques such as noise injection, pixelation, and pruning to these layers rather than the whole model, minimizing the attack surface and preserving client data privacy. We evaluate \DRArmor's performance against the advanced LoKI attack across diverse datasets, including MNIST, CIFAR-10, CIFAR-100, and ImageNet, in a 200-client FL setup. Our results demonstrate \DRArmor's effectiveness in mitigating data leakage, achieving high True Positive and True Negative Rates of 0.910 and 0.890, respectively. Additionally, \DRArmor maintains an average accuracy of 87\%, effectively protecting client privacy without compromising model performance. Compared to existing defense mechanisms, \DRArmor reduces the data leakage rate by 62.5\% with datasets containing 500 samples per client. 
\end{abstract}



\maketitle
\section{Introduction}
\label{sec:intro}

In Federated Learning (FL) \cite{mcmahan2017communication}, data is not centralized for training and remains distributed between devices in the network. This ensures that raw user data never leaves the local devices, significantly reducing the risk of exposure to external threats. By sharing only model updates rather than user device data, FL mitigates potential privacy concerns while enabling collaborative model improvement. \par 

The promise of FL to ensure user data privacy is contingent upon the security of the gradients sent from the client nodes to the server. Studies \cite{zhao2024LoKI} have shown that gradients sent from client nodes can reveal properties of the underlying client data or the data itself. Property inference \cite{luo2021feature,melis2019exploiting}, membership inference \cite{choquette2021label,nasr2019comprehensive,shokri2017membership}, and GAN-based attacks \cite{hitaj2017deep,wang2019beyond} have successfully inferred data stored in client nodes. A more severe class of attacks, as described in \cite{zhao2024LoKI}, involves a malicious server that tries to steal private client training data. These types of attacks fall into two categories -- Optimization attack and analytical attacks. While optimization attacks \cite{yin2021see,zhao2024LoKI}, which typically target image data, have been mitigated by secure aggregation methods in the FL architecture, analytic reconstruction attacks \cite{boenisch2023curious,fowl2021robbing} remain a threat. In optimization attacks, a random dummy sample is initialized and then the difference between the true gradient and the generated one is optimized. However, the quality of reconstructions degrades with the increasing size of batch size \cite{yin2021see}. On the other hand, analytical attacks customize model parameters or model architecture to extract training data from the model layers. \par
One of the most notable Data Reconstruction Attacks (DRA) is the ``Robbing the Fed'' (RtF) attack \cite{fowl2021robbing}, which demonstrates how minimal but malicious modifications to the shared model architecture can enable the server to directly obtain a verbatim copy of user data from gradient updates. This attack is particularly concerning because it bypasses the need for solving complex inverse problems, making it highly efficient and scalable. The implications of such an attack can be severe, as it can compromise the privacy of user data even when aggregated over large batches. A more recent attack, LoKI, is proposed by authors in \cite{zhao2024LoKI}. LoKI overcomes the limitations of RtF by targeting the FedAVG setting and using customized convolutional parameters to maintain the separation of weight gradients between clients, even through aggregation. This attack has shown the ability to leak a substantial portion of training data, with success rates of 76–86\% in a single training round, even when secure aggregation is employed. The effectiveness of LoKI further highlights the vulnerability of FL systems to sophisticated model manipulation techniques.\par

An increasing number of solutions have emerged to protect the privacy of data in FL. Secure aggregation methods \cite{chen2024federated,qi2019visualizing} have been proposed to mitigate such threats. However, they remain vulnerable to advanced attacks such as LoKI. Byzantine-resilient aggregation techniques \cite{so2020byzantine,zhao2021sear} can identify and exclude outlier gradients that deviate significantly from the majority, thereby reducing the risk of data leakage. However, in these methods, the server can manipulate the weight gradients to a normalized value, so that the gradients are not detected. Anomaly detection algorithms \cite{mothukuri2021federated} can monitor the training process for unusual patterns indicative of an attack. However, skilled attackers can design their malicious updates to evade detection by anomaly detection algorithms, reducing their effectiveness. Additionally, differential privacy (DP) \cite{jayaraman2019evaluating,kim2021survey,wei2020federated} involves adding noise to the model updates before they are sent to the server, which can prevent a server from fully reconstructing private data. However, this method significantly decreases model performance, especially with large vision models \cite{zhang2022understanding}. As mentioned in \cite{zhao2024LoKI}, because the attack could occur at any point in the training process, differential privacy would need to be applied at every training step, making it extremely costly in terms of accuracy. \par

The current solutions fall short against sophisticated attacks without compromising the utility of FL models. Existing approaches predominantly treat models as black-box systems, limiting their ability to address the underlying vulnerabilities. The authors in \cite{zhao2024LoKI} emphasize that identifying modified layers in a model is particularly challenging due to their potential to be embedded in various locations within a larger architecture. This challenge arises because current solutions do not analyze the inner workings of the model's layers -- the very components that might facilitate DRA. To bridge this gap, we propose a novel methodology that treats the model as \textit{a white-box system by focusing on the layers involved in gradient updates}. By leveraging Explainable AI (XAI), our approach seeks to understand how these layers contribute to the model's output and identify those whose gradients deviate from normal behavior, indicating an intent to reconstruct data rather than improve the model's performance. Further, most related research \cite{rodriguez2023survey} focuses on server-side mechanisms or aggregation methods to safeguard the data of client nodes. In this work, our goal is to \textit{identify the root cause of the attack on the client side}, irrespective of the aggregation algorithms on the server side. To the best of our knowledge, this is the first work to take advantage of explainable AI to design a robust algorithm to detect malicious layers in a modified model architecture aimed at reconstructing client data in FL. \par

\textbf{Our work.} We introduce \DRArmor, a novel system designed to detect and mitigate malicious layers in DRA-modified models, such as those introduced in Rtf \cite{fowl2021robbing} and LoKI \cite{zhao2024LoKI}. By leveraging XAI techniques, including Layer-wise Relevance Propagation (LRP) \cite{bach2015pixel} and Deep Taylor Decomposition (DTD) \cite{montavon2017explaining} on these replicated attacks, \DRArmor explains how each layer of a model contributes to its output. Layers exhibiting low relevance but high learning—a characteristic often linked to malicious intent—are flagged by the system. These layers are then addressed and mitigated, ensuring only the truly relevant layers remain in the model. This process significantly enhances the security of client devices in FL environments. Unlike traditional approaches that rely heavily on centralized server-side defenses, \DRArmor enables clients to take proactive control by detecting and responding to DRA attacks independently. This shift to client-side protection addresses the root cause of such attacks more effectively. Our evaluation demonstrates the system's effectiveness, achieving True Positive Rates (TPR) of 0.910 and True Negative Rates (TNR) of 0.890 across multiple datasets. When malicious layers are identified, \DRArmor employs targeted techniques such as adding DP noise or pixelated noise solely to these layers. This approach is more effective than applying standard DP methods (e.g., \cite{kim2021survey}) across the model. Additionally, \DRArmor supports pruning irrelevant layers along with preserving model performance. The system's robustness is further highlighted by its ability to maintain high detection accuracy and significantly reduce data leakage, even as the number of client nodes and dataset sizes increase. Moreover, \DRArmor is effective against both continuous and periodic poisoning attacks, ensuring that model accuracy remains intact with minimal degradation. \par

Our main contributions are: 
\begin{enumerate}
    \item We introduce \DRArmor, a novel defense solution that uses XAI to detect and mitigate malicious layers in DRA-modified models. Even with extended replications of attacks such as LoKI with 200 clients and dataset sizes of 500 per round, \DRArmor achieved 87\% detection accuracy, including cases where malicious layers were deeply embedded in the model architecture.
    \item Unlike traditional server-side methods or aggregation defenses on the entire model, \DRArmor empowers client devices to identify and defend against malicious layers independently. \DRArmor applies targeted defenses, such as DP noise and pixelation, exclusively to malicious layers, leaving the rest of the model unaffected.
    \item \DRArmor consistently detects and mitigates DRA across iterations, demonstrating robustness against both continuous and periodic poisoning attacks while maintaining model accuracy.
    \item Compared to existing defense mechanisms, \DRArmor reduces the data leakage rate by 62.5\% on networks with 200 clients and datasets of 500 samples per client.
\end{enumerate}

\section{Background and Related Work}
\label{sec:relwork}

\subsection{DRA on FL}
While numerous attacks on FL have been proposed, our focus is on defenses against DRA, in which the model architecture is altered to directly obtain client data through their updates. These attacks exploit vulnerabilities in the gradient calculation and sharing processes in FL. \par
Optimisation-based attacks \cite{yin2021see,dimitrov2022data} operate by iteratively refining a dummy data sample to match the gradient of the actual training data. 

Although these attacks are effective with small batch sizes, they suffer from decreased quality and require more iterations as batch sizes increase. They often fail with larger image resolutions and aggregated gradients, and they are largely limited to FedSGD \cite{mcmahan2017communication}. Some of these methods assume knowledge of the user labels, which are obtained outside of the optimisation process. Additionally, optimisation - based attacks can produce reconstructed images that display characteristics of the image class but are not actually present in the training image, as shown in \cite{yin2021see}.\par
Different methods have been proposed targeting FL with secure aggregation, often with limited success. One approach \cite{wen2022fishing} is to magnify gradients to target a single data point. This involves manipulating weights so that the gradients of a targeted class are magnified by reducing model confidence in that class’s prediction. Another approach \cite{pasquini2022eluding} aims to make the aggregated gradient the same as an individual client's gradient. This can be achieved by sending model parameters that result in zero gradients for all but the targeted client. There are also methods \cite{kariyappa2023cocktail} that treat the inputs of an FC layer as a blind source separation problem. However, these methods are often limited in scale, typically only managing to extract data from a single user per training round or having restrictions on the number of inputs.\par
Analytic attacks, particularly those involving linear layer leakage \cite{fowl2021robbing,boenisch2023curious}, directly extract training data from the gradients of FC layers. These attacks modify model parameters or architecture to retrieve inputs from the weight and bias gradients of an FC layer. When only one data sample activates a neuron in an FC layer, the input can be directly computed from the gradients. These methods can reconstruct inputs with high accuracy but typically fail when multiple data samples activate the same neuron. As shown in \cite{zhao2024LoKI}, they also face scalability issues when dealing with larger batch sizes or secure aggregation, which conceals individual client updates. While increasing the size of the FC layer can maintain a high leakage rate, it often results in very large models. LoKI improves DRAs by introducing techniques to overcome scalability challenges and secure aggregation limitations. Unlike the previous methods, it enables simultaneous data reconstruction from multiple clients in a single training round, addressing the inefficiencies of optimisation-based and analytic attacks. Key features of LoKI include the use of customized convolutional kernels for each client, allowing the server to separate and recover client-specific data even after aggregation. This attack highlights critical vulnerabilities in FL systems and motivates the development of robust defenses.

\subsection{Existing Defense Mechanisms}

\textbf{Differential Privacy (DP).} DP \cite{dwork2006differential} is a privacy-preserving technique that introduces random noise to the data or model updates to protect individual data points from being inferred. In the context of FL, DP can be applied by adding noise to the gradients before they are shared with the server as surveyed in \cite{el2022differential}. This noise addition ensures that the contribution of any single data point is obscured, making it difficult for an attacker to reconstruct the original data. \par

Mathematically, differential privacy guarantees for any two datasets $D$ and $D'$ differing by a single data point, the probability of obtaining a particular output $O$ is nearly the same. This can be expressed as:
\(
\Pr[\mathcal{M}(D) = O] \leq e^{\epsilon} \cdot \Pr[\mathcal{M}(D') = O],
\)
where \(\mathcal{M}\) is the randomized mechanism (e.g., the gradient update process), and \(\epsilon\) is the privacy budget that controls the level of privacy. A smaller \(\epsilon\) indicates stronger privacy but may introduce more noise.\par
In the context of DRA, the added noise affects the reconstructed data by making the gradients less precise. Suppose the true gradient is \(g\), and the noise added is drawn from a Laplace distribution with scale parameter \(\frac{\Delta g}{\epsilon}\), where \(\Delta g\) is the sensitivity of the gradient. The noisy gradient \(g'\) can be represented as:
\(
g' = g + \text{Lap}\left(\frac{\Delta g}{\epsilon}\right).
\)

The effectiveness of DP in preventing DRA is due to the noise, which makes it challenging for the attacker to accurately infer the true data - especially when the privacy \(\epsilon\) is small. Mironov \cite{mironov2017renyi} used Rényi divergence for more accurate privacy loss calculations, and Abadi et al. \cite{abadi2016deep} created DP-SGD, an empirical algorithm that applies DP to machine learning training. Other research \cite{agarwal2018cpsgd,andrew2021differentially,bu2024automatic,xia2023differentially} builds upon DP-SGD by improving its performance through methods such as adaptive gradient clipping and suitable perturbations. Authors in \cite{levy2021learning} also apply DP to FL to protect user-level privacy and ensuring data privacy when collecting local parameters. However, this comes at the cost of potentially degrading the model's performance, highlighting the trade-off between privacy and utility as shown in \cite{cormode2011personal,dick2023confidence}.\par

 Gradient compression is also commonly adopted by studies like \cite{yin2021see} as a baseline defense to examine the robustness of attacks. Recently, Soteria \cite{sun2021soteria} leveraged a similar idea to gradient compression but with a smarter pruning strategy to decrease privacy leakage risks with similar performance. Recent works such as \cite{balunovic2022lamp} evaluated these defense methods and broke them by introducing a Bayesian optimal adversary. The Bayesian framework provides a way to analyse the problem of gradient leakage by considering the joint probability distribution of inputs and their gradients. This framework allows for the formulation of a Bayes optimal adversary as an optimisation problem, which minimises the risk of reconstructing the input given the observed gradients. The paper demonstrates that existing attacks can be interpreted as approximations of this optimal adversary, each making different assumptions about the underlying probability distributions. This theoretical grounding enables the development of stronger attacks by leveraging the knowledge of these distributions, proving more effective than previous methods.\par
 
\textbf{Limiting Mutual Information (MI).} Several studies \cite{barthe2011information,cuff2016differential} explore the connection between MI and DP by deriving upper bounds of MI for different DP mechanisms. Other studies \cite{li2020tiprdc,osia2018deep} propose training feature extractors that reduce the MI between the output features and the assigned labels while maximising the MI between the output features and the original data. Fisher information has also been used to measure information leakage, aligning with the theories in \cite{hannun2021measuring}. Authors in \cite{uddin2020mutual} model FL using information theory to measure MI between variables in FL. A more recent study \cite{tan2024defending} proves that the reconstruction error of DRA is directly related to the information an attacker gains through shared parameters, as measured by MI. To limit this information leakage, the method establishes a channel model where the shared parameters act as a communication channel and then limits the channel capacity by adding Gaussian noise to the parameters. This effectively reduces the information available to an attacker, making data reconstruction challenging. \par

The defense methods reviewed may effectively limit the information shared by client devices with the server. However, they do not address one fundamental issue: \textit{client devices cannot detect how or where gradients are being leaked}. Moreover, these methods lack transparency in explaining their effectiveness. We postulate that if the nature of the attack evolves, these approaches may fail in detection. As mentioned by the authors of LoKI, since the malicious layers can be placed at any part of the model architecture, it would be challenging to detect such attacks if the attacker modifies the parameters. Additionally, our approach offers computational savings as the defense mechanism is applied only to the malicious layers rather than the entire model.

\subsection{Preliminary - Explainable AI}
 
Explainable AI (XAI) refers to techniques used to provide insights into how models arrive at their predictions, enabling users to understand, trust, and effectively manage AI systems. It helps identify biases and improve model performance. There are multiple varieties of XAI methods. In this work, we are using the following:\par

\textbf{Layer-wise Relevance Propagation (LRP).} LRP is a technique used to explain the predictions of deep neural networks by propagating the prediction backward through the network. LRP assigns relevance scores to each input feature, indicating their contribution to the final prediction. Mathematically, the relevance score \( R_j \) for a neuron \( j \) in layer \( l \) is computed by distributing the relevance from the neurons in the subsequent layer \( l+1 \):
\begin{equation}
R_i^{(l)} = \sum_j \frac{z_{ij}}{\sum_{i'} z_{i'j} + \epsilon \cdot \mathrm{sign}(\sum_{i'} z_{i'j})} R_j^{(l+1)}
\label{eq:lrp}
\end{equation}

   where
    $z_{ij}$ is the contribution of neuron $i$ in layer $l$ to neuron $j$ in layer $l+1$,
    $\epsilon$ is a small stabilization term to prevent division by zero, and 
    $R_j^{(l+1)}$ is the relevance of neuron $j$ in the next layer $l+1$. LRP helps in visualizing which parts of the input data are most relevant to the model's output, making it easier to interpret complex neural networks and identify potential issues. \par
\textbf{Deep Taylor Decomposition (DTD).} DTD is an advanced method for explaining the predictions of deep neural networks by decomposing the prediction into contributions from each input feature. DTD is based on Taylor series expansion, which approximates a function using its derivatives. Mathematically, the relevance \( R_i \) of an input feature \( x_i \) is computed by decomposing the relevance of the output \( R \) into contributions from the input features:
   \(
   R_i = x_i \frac{\partial R}{\partial x_i}.
   \)
   By iteratively applying this decomposition from the output layer to the input layer, DTD provides a detailed explanation of the model's decision-making process, highlighting the importance of each input feature.

\section{System and Threat Model}
\label{sec:problem_statement}
\textbf{System Model.}
The components of the FL-based system used in this work include Server, Client Device, and Attacker.
The Server is responsible for registering a model, iteratively sending it to the clients, aggregating the client updates, and training the global model. The Client Device is responsible for storing the data, training the model received from the server, and sending updates back to the server. The Attacker is responsible for altering model parameters and architecture to obtain updates from the client, with the goal of reconstructing the client's data.
The system consists of one server, multiple client devices, and multiple attackers. The client devices communicate directly with the server and train the model received from the server. We consider a real-life FL scenario in which any device can receive any model, and the client cannot assume that a model received in the iteration after a malicious one will also be malicious. 
Unlike the studies in \cite{zhao2024LoKI} and \cite{fowl2021robbing}, which used 100 clients, our experiments scale up to 200 clients to evaluate the impact of these attacks on model accuracy in a larger setting. \par

\textbf{Threat Model.}

We consider a standard FL setting where a central server coordinates model training across multiple clients, each holding private local data. The server initializes and distributes a global model, while clients perform local training and return gradient updates for aggregation. In our threat model—adopted from LoKI~\cite{zhao2024LoKI} and originally introduced in~\cite{fowl2021robbing}—the server follows the FL protocol but may attempt to reconstruct private client data by modifying the model architecture before the distribution. Specifically, the server can inject malicious layers into the model to exploit gradient leakage for DRA. However, the server cannot deviate from the standard federated learning protocol in ways beyond changes to model architecture or parameters, and such modifications must remain within the operational limits of standard machine learning frameworks. Clients are assumed to be benign, unaware of these manipulations, and follow the protocol by training the received model on local data and returning gradients. The adversarial model includes a single attacker—the server—with no client-side collusion.

\section{\DRArmor}
\label{sec:methodology}
\DRArmor is designed to address the critical challenge of detecting and mitigating malicious layers introduced by DRA in FL systems. By leveraging Explainable AI (XAI), \DRArmor analyzes the model architecture as a white-box, examining the model architecture and the weights of each layer to detect any modifications indicative of malicious intent. \par

As described in \cite{zhao2024LoKI,fowl2021robbing}, the input \( x_i \) is calculated by dividing the weight gradient by the bias gradient, as shown in the equation:  
\(
x_i = \frac{\delta L / \delta W_i}{\delta L / \delta B_i},
\)
where \( i \) is the activated neuron, and \( \frac{\delta L}{\delta W_i} \) and \( \frac{\delta L}{\delta B_i} \) are the weight and bias gradients of the neuron, respectively. In the context of detecting malicious layers, the variables \( \frac{\delta L}{\delta W_i} \) and \( \frac{\delta L}{\delta B_i} \) play a critical role. Malicious layers, designed to reconstruct private data rather than contribute meaningfully to the overall model performance, exhibit gradients that are not aligned with the model's intended learning objectives. Specifically, the weight gradients \( \frac{\delta L}{\delta W_i} \) in these layers reflect manipulation aimed at data leakage, leading to anomalous behavior such as higher magnitude or misalignment with gradients of other layers. These manipulated gradients fail to contribute to improving the model's predictive performance, making them indicative of malicious intent. This often indicates that the model is trying to extract information that does not significantly contribute to the final output. This is where Explainable AI (XAI) becomes crucial. XAI techniques help us understand which neural network layers are actually contributing to the model's output. For instance, if a model is designed to classify an image as a cat, certain layers focusing on reconstructing the data might not directly contribute to the final classification of ``cat''. These layers might produce higher gradients than those more relevant to the classification task. By using XAI, we can identify and understand the roles of different layers in the model, ensuring that the model's decision-making process is transparent and interpretable.\par
Two of the most popular methods for understanding the relevance of a layer to the output are Layer-wise Relevance Propagation (LRP) and Deep Taylor Decomposition (DTD). Our method leverages the explainability provided by LRP and DTD to identify layers that contribute disproportionately to the model's output, which indicates potential malicious behavior.

Suppose we have a neural network model designed to classify images along with the additional malicious layers described in \cite{zhao2024LoKI}, and we input an image of a cat. These layers might perform a DRA, making them irrelevant to the final classification task. 
The standard propagation rule for relevance in LRP is given by Equation~\ref{eq:lrp}.

To tailor LRP to our needs, we modify the propagation rule to emphasize the discrepancy between the intended contributions of layers and their actual gradient behaviors. Specifically, for a layer $l$, we adapt the rule as:
\(
R_i^{(l)} = \sum_j \frac{z_{ij}}{\sum_{i'} z_{i'j} + \epsilon} \cdot \gamma \cdot R_j^{(l+1)},
\)
where $\gamma \in [0,1]$ is a scaling factor that adjusts the contribution of relevance attribution of the previous layer. It is computed adaptively to reflect the model's certainty about the importance of a given layer. Layers that exhibit high gradients and are deemed highly relevant contribute more strongly to the propagation through a higher \( \gamma \), while less relevant or low-confidence layers result in a smaller \( \gamma \). This dynamic adjustment ensures that the propagation process emphasizes layers that are both significant and exhibit confident behavior in terms of relevance attribution. 

\subsection{Detecting Malicious Layers}
In a benign model, the layers collaborate to generate the output by focusing on relevant input features. 

Malicious layers are modified to reconstruct input data rather than contributing to output generation. This discrepancy is manifested in their gradients and relevance scores:\\
\textbullet\ \textbf{Relevance Scores:} Malicious layers contribute minimally to the output relevance. When propagating relevance using the modified LRP rule, these layers will exhibit low relevance scores ($R^{(l)} \approx 0$) since their primary purpose is not aligned with the model's intended task.\\
\textbullet\ \textbf{Gradient Analysis:} Malicious layers are designed to reconstruct input data, leading to unusually high gradients with respect to input features. Mathematically, the gradient $\nabla_{\mathbf{x}} f^{(l)}$ of the malicious layers will exhibit large magnitudes, indicating their strong dependence on the input data for reconstruction. Gradient analysis is incorporated by evaluating the norm of gradients $\nabla_{\mathbf{x}} f^{(l)}$ for each layer:
\(
G^{(l)} = \| \nabla_{\mathbf{x}} f^{(l)} \|,
\)
where $G^{(l)}$ captures the gradient of the layer with respect to input features for each layer $l$. Layers with anomalously high gradients with low relevance are flagged as potentially malicious, as they disproportionately influence data reconstruction. \par

\textbf{Gradient-Relevance Discrepancy.}
To formally identify malicious layers, we define a metric based on the gradient relevance discrepancy. For each layer $l$, we compute:
\(
D^{(l)} = \frac{\| \nabla_{\mathbf{x}} f^{(l)} \|}{\| R^{(l)} \|},
\)
where $\| \cdot \|$ denotes the norm (e.g., $L_2$ norm).

A high value of $D^{(l)}$ indicates a layer with large gradients but low relevance, characteristic of malicious behavior. Using a threshold $\tau$, we classify a layer $l$ as malicious if:
\(
D^{(l)} > \tau.
\)

\textbf{Justification for LRP Modifications.}
The modifications to the standard LRP formulation are critical for adapting it to the detection of malicious layers. By introducing the scaling factor $\gamma$, we ensure that the propagation process captures subtle but critical deviations in layer behavior. For example, benign layers that contribute to the output will maintain proportional relevance scores under this adjustment, while malicious layers will show diminished relevance.\par
Furthermore, by computing the discrepancy $D^{(l)}$, we leverage the complementary information provided by gradients and relevance. Gradients highlight sensitivity to input changes, while relevance scores reflect actual contribution to the output. Malicious layers, which prioritize data reconstruction over task alignment, are naturally distinguished by this divergence (see \S\ref{subsec:detect_mal_layers}).\par
While LRP is effective for small and moderately sized models, its performance degrades with deeper architectures and models with large numbers of parameters. These limitations arise due to:\\
\textbullet\ \textbf{Vanishing Relevance:} As relevance is propagated backward through multiple layers, it may diminish or concentrate disproportionately in certain neurons, making it challenging to identify malicious layers accurately.\\
\textbullet\ \textbf{Sensitivity to Stabilization Terms:} The stabilization term $\epsilon$ can significantly impact the relevance propagation process, leading to inconsistent results for different architectures.\\
\textbullet\ \textbf{Computational Overhead:} For large-scale models, the backward relevance propagation requires substantial computational resources, making it computationally expensive for large-scale analysis (see Appendix \ref{subsec:scalability}). \par
To address these limitations, we propose using DTD as an alternative for large-scale models. DTD provides a more robust mechanism for layer-wise relevance analysis by decomposing the model's output into contributions from each layer using a Taylor series expansion (see \S\ref{subsec:detect_mal_layers}). The incorporation of Wasserstein distance further refines this process by quantifying the similarity between reconstructed relevance distributions and expected distributions.\par
For a neural network $f(\mathbf{x})$ with output $y$, DTD approximates the relevance of layer $l$ by considering the Taylor expansion of $f$ around a root point $\mathbf{x}_0$ (typically chosen to minimise$f$):
\[
R^{(l)} = f(\mathbf{x}) - f(\mathbf{x}_0) \approx \nabla_{\mathbf{x}} f^{(l)} \cdot (\mathbf{x} - \mathbf{x}_0).
\] \par
Unlike LRP, DTD derives relevance scores directly from the gradient information relative to an input perturbation, making it more robust against the vanishing relevance issue that affects LRP. The term $(\mathbf{x} - \mathbf{x}_0)$ ensures that the decomposition aligns with localized input changes, capturing the contributions of malicious layers more effectively. A key advantage of DTD is that it avoids the dependency on stabilization terms ($\epsilon$) that can introduce inconsistencies in LRP. Mathematically, DTD provides a more accurate decomposition by considering higher-order derivatives implicitly, ensuring that the relevance scores remain consistent across deep architectures.\par

\textbf{Wasserstein Distance for Consecutive Layers.}
The Wasserstein distance is a metric for comparing probability distributions. In this context, we use it to measure the similarity between relevance distributions of consecutive layers $R^{(l)}$ and $R^{(l+1)}$. Mathematically, the Wasserstein distance is defined as:
\[
W(R^{(l)}, R^{(l+1)}) = \inf_{\gamma \in \Gamma(R^{(l)}, R^{(l+1)})} \int \|r - r'\| \; d\gamma(r, r'),
\]
where $\Gamma(R^{(l)}, R^{(l+1)})$ is the set of all joint distributions with marginals $R^{(l)}$ and $R^{(l+1)}$. A high Wasserstein distance between consecutive layers indicates a significant deviation in their relevance behavior, flagging the layer as potentially malicious.

By applying Wasserstein distance to consecutive layers, we capture the transition irregularities introduced by malicious layers. This approach avoids reliance on predefined reference distributions. \par

The defense methodology using DTD and Wasserstein distance is implemented as follows:
\begin{enumerate}
    \item Perform DTD to compute relevance scores $R^{(l)}$ for all layers based on their Taylor expansion contributions.
    \item Calculate the Wasserstein distance $W(R^{(l)}, R^{(l+1)})$ for each pair of consecutive layers.
    \item Compute the gradient-relevance discrepancy $D^{(l)}$ for additional validation:
    \(
    D^{(l)} = \frac{\| \nabla_{\mathbf{x}} f^{(l)} \|}{\| R^{(l)} \|}.
    \)
    \item Flag layers with $W(R^{(l)}, R^{(l+1)}) > \tau_W$ or $D^{(l)} > \tau_D$ as malicious.
    \item Retrain or replace flagged layers to mitigate their impact on the model.
\end{enumerate}
DTD, as used in DRArmor, is gradient-based and avoids recursive backpasses, making it significantly more efficient and well-suited for typical federated clients with limited resources. Similarly, the Wasserstein distance computation is applied once per round for layer-wise distributions and can be computed in linear time with entropic regularization. These components are executed once per local round (not per batch or epoch), thus incur minimal additional overhead relative to standard local training. 
\subsection{Defense against Malicious Layers}
Once malicious layers are identified within the model, the next critical step involves implementing appropriate defensive actions to prevent DRA. The following strategies can be employed as mitigation techniques:\par
    \textbf{Differential Privacy-Based Noise Addition.} Differential Privacy (DP) is a widely recognized strategy to mitigate information leakage in FL. While the authors of LoKI suggest that applying DP universally by adding noise to all gradient updates can significantly reduce model accuracy, a more targeted approach can strike a balance between privacy and performance. Instead of adding noise to all gradient updates, noise is injected specifically into the updates originating from the malicious layers. This targeted application minimises the impact on overall model accuracy while effectively mitigating the DRA. Since only the malicious layers are affected, the integrity of legitimate layers remains intact, preserving the utility of the model. Moreover, noise levels can be dynamically adjusted based on the confidence in the detection of malicious layers. For instance, layers identified with higher certainty can have more noise added, while those with marginal detection can have less aggressive noise injection.\par
 Let \( \mathbf{g} = [g_1, g_2, \dots, g_L] \) represent the gradients of a model with \( L \) layers, where \( g_l \) is the gradient vector of the \( l \)-th layer. Assume \( \mathcal{M} \subset \{1, 2, \dots, L\} \) denotes the set of malicious layers identified by the detection mechanism. For each \( l \in \mathcal{M} \), noise is added as follows:
\(
\tilde{g}_l = g_l + \mathcal{N}(0, \sigma^2),
\)
where \( \mathcal{N}(0, \sigma^2) \) is Gaussian noise with mean \( 0 \) and variance \( \sigma^2 \). Layers not identified as malicious remain unchanged, i.e., \( \tilde{g}_l = g_l \) for \( l \notin \mathcal{M} \).\par
The noise level \( \sigma^2 \) can be dynamically adjusted based on the confidence score \( c_l \) for identifying layer \( l \) as malicious:
\(
\sigma^2_l = \sigma^2_{\text{base}} \cdot (1 + \alpha \cdot c_l),
\)
where 
    \( \sigma^2_{\text{base}} \) is base noise level, 
    \( c_l \in [0, 1] \) is confidence score for layer \( l \) being malicious, 
    and \( \alpha \) is scaling factor that controls the impact of confidence on noise level.

Higher confidence in detecting a malicious layer results in larger noise being added, ensuring better obfuscation of sensitive information.\par
After noise injection, the updated gradients \( \tilde{\mathbf{g}} \) are aggregated and sent to the server:
\(
\tilde{\mathbf{g}} = [\tilde{g}_1, \dots, \tilde{g}_L].
\)
This ensures that gradients from malicious layers are obfuscated while preserving the integrity of legitimate layer gradients.\par

Targeted DP minimizes the degradation of model performance. The change in accuracy due to noise application, denoted as \( \Delta_{\text{acc}} \), is given by:
\(
\Delta_{\text{acc}} = \text{Accuracy}_{\text{no DP}} - \text{Accuracy}_{\text{DP}}.
\)
Since only malicious layers are affected, \( \Delta_{\text{acc}} \) remains significantly smaller compared to universal DP application. (see \S\ref{subsec:defence})\par

\textbf{Pixelating gradients.} Pixelation involves reducing the gradient by grouping adjacent values into blocks and replacing them with a single representative value, such as the average or median intensity of the block. By applying pixelation to the gradients sent from malicious layers, the precision of the gradient information is reduced, thereby limiting the ability of the server to reconstruct the original client data.\par
Given a gradient matrix \( G \) of dimensions \( m \times n \), the pixelation process operates as follows:\par
The gradient matrix \( G \) is divided into non-overlapping blocks of size \( b \times b \). The total number of blocks is given by:
\(
N = \frac{m}{b} \times \frac{n}{b},
\)
where \( b \) is the block size, chosen such that \( m \) and \( n \) are divisible by \( b \). Each block \( B_{ij} \) corresponds to a sub-matrix:
\(
B_{ij} = G[i \cdot b:(i+1) \cdot b, j \cdot b:(j+1) \cdot b],
\)
where \( i \in \{0, \ldots, \frac{m}{b} - 1\} \) and \( j \in \{0, \ldots, \frac{n}{b} - 1\} \).\par
Each block \( B_{ij} \) is replaced by the representative mean:
\(
\hat{B}_{ij} = \text{mean}(B_{ij}) = \frac{1}{b^2} \sum_{p=1}^{b} \sum_{q=1}^{b} G_{p,q}.
\).
The pixelated gradient matrix \( \hat{G} \) is constructed by replacing each block \( B_{ij} \) in \( G \) with its corresponding representative value \( \hat{B}_{ij} \):
\(
\hat{G}[i \cdot b:(i+1) \cdot b, j \cdot b:(j+1) \cdot b] = \hat{B}_{ij}.
\) \par
 By pixelating gradients \( G \) from malicious layers, the client can obfuscate sensitive information, reducing the precision of data reconstruction attempts. The server’s ability to reconstruct client data is typically dependent on 
\(
 G,
\) which
 with pixelated gradients, becomes \( \hat{G} \).
The quantization effect introduced by pixelation disrupts the reconstruction process, protecting client data while retaining the model's learning capacity (see \S\ref{subsec:defence}).
\par
    \textbf{Pruning Malicious Layers.} Another effective strategy involves pruning the identified malicious layers entirely from the model architecture. This approach relies on bypassing the malicious layers while maintaining the flow of relevant information through the network. After pruning a malicious layer, a skip connection can be introduced to route the input image (or feature map) directly to the next relevant layers. Ensuring that the dimensions match is crucial for the network to function correctly when implementing skip connections. If they do not match, we can use techniques like padding, cropping, or applying a linear transformation to adjust the dimensions. In our experiment, we use a linear layer along with padding to transform the input to the required dimensions. Consider an input image of size $(64 \times 64 \times 3)$ and a target convolutional layer (non-malicious) with a filter size of $(3 \times 3)$ and 32 filters as an example. First, we apply an initial convolutional layer with the same filter size and number of filters to the input image, using a stride of 1 and padding to maintain the spatial dimensions. This results in an output size of $(64 \times 64 \times 32)$. Next, we bypass the potentially malicious layer by directly connecting the adjusted input to the third convolutional layer. To ensure dimensional compatibility, we use a convolutional layer with the same filter size and number of filters as the target layer, resulting in an adjusted output size of $(64 \times 64 \times 32)$. This preprocessing step ensures that the input image can be effectively integrated into the network at the desired point, preserving the data's integrity and enhancing the network's robustness.

\section{Experimental Setup}
\label{sec:experiments}

The experiments were carried out in a simulated FL environment, where we implemented the defense system components using the Keras and Flower \cite{beutel2020flower} frameworks. The system was executed on a server equipped with an NVIDIA H1000NVL GPU, an Intel Xeon 2.10 GHz CPU with 8 cores, and 96 GB of RAM. The system's scalability was validated by simulating 200 clients, a scenario representative of real-world FL environments. We selected widely used datasets for our experiments, including MNIST, CIFAR-10, CIFAR-100, ImageNet, and Cats vs Dogs. These datasets were chosen because they are commonly used in related research for evaluating FL systems and defense mechanisms, offering a benchmark for comparison. Additionally, their diverse characteristics -- ranging from simple grayscale digits to complex natural images -- enable us to thoroughly assess the effectiveness of our approach across various data types and scenarios. The associated model configurations for each dataset are detailed below.\par
    \textbf{MNIST} \cite{deng2012mnist}. This is a set of 60,000 handwritten numbers from 0--9. Each data consists of $28\times 28$ pixel grayscale images. We used a CNN model with three convolutional layers, where the first layer consists of 32 output channels with kernel size 3 and stride 1. The second convolutional layer outputs 64 channels with the same kernel size and stride of 3 and 1, respectively. The third convolutional layer outputs 256 channels with the same kernel size and stride of 3. It is followed by two dropout layers and two fully connected layers. The final output is given by a log softmax layer that predicts the number denoted in the input image.\par
    \textbf{CIFAR-100} \cite{krizhevsky2009learning}. This dataset consists of 60,000 $32\times 32$ color images in 100 classes, with 600 images per class. There are 50,000 training images and 10,000 test images. The model begins with an initial convolutional layer with 64 output channels, a kernel size of 3, and a stride of 1, followed by batch normalization and a ReLU activation. Subsequent layers consist of repeated residual blocks, each comprising two convolutional layers with 64, 128, and 256 output channels as the depth increases. Each convolution is followed by batch normalization and ReLU activation. To handle overfitting, dropout layers are interspersed between some convolutional blocks. After the convolutional and residual layers, global average pooling is applied to reduce the spatial dimensions. The fully connected layers include one hidden layer of 512 units followed by a final output layer with 10 units corresponding to the CIFAR-10 classes. The final predictions are produced using a log softmax layer.\par
    \textbf{CIFAR-10} \cite{krizhevsky2009learning}. This dataset consists of 60,000 $32\times 32$ color images in 10 classes, with 6,000 images per class. There are 50,000 training images and 10,000 test images. We used the same model as used for CIFAR-10.\par
    \textbf{ImageNet} \cite{deng2009imagenet}. This dataset contains over 14 million images across 1,000 classes. For our experiments, we used a subset of 1.2 million training images and 50,000 validation images. The model begins with an initial convolutional layer that outputs 64 channels, using a kernel size of 7 and a stride of 2, followed by batch normalization, ReLU activation, and max pooling to reduce spatial dimensions. The subsequent layers are structured as residual blocks. Each block contains two convolutional layers with increasing output channels -- 64, 128, 256, and 512 -- as the network depth increases. All convolutions are followed by batch normalization and ReLU activation, with downsampling implemented through strided convolutions. To prevent overfitting and improve generalization, dropout layers are interspersed between some residual blocks. The feature maps are then reduced using global average pooling before being passed through two fully connected layers. The final output layer consists of 1,000 units, corresponding to the ImageNet classes, and predictions are generated using a log softmax function.\par
    \textbf{Cats v Dogs} \cite{parkhi2012cats}. For this dataset, we employed the widely-used ResNet-18 architecture, which balances computational efficiency and performance.\par
Each dataset was divided into training, testing, and validation sets, with a sampling ratio of 60\%, 30\%, and 10\%, respectively. The primary objective of our experiments was to detect and defend against malicious layers attempting to execute DRA within a system of 200 clients -- twice the number used in the LoKI attack. All models were modified by adding a convolutional layer and two FC layers, initiating with the start of the model and then proceeding further into the model architecture. Additionally, we evaluated the behavior of \DRArmor under varying numbers of layers, with the results presented in \S\ref{sec:eval}. In all experimental settings, we used a learning rate of $\alpha = 1 \times 10^{-4}$. Local models were trained for 15 to 200 rounds, depending on the complexity of the model architecture and the size of the dataset. The training process aimed to achieve stable accuracy while avoiding overfitting, ensuring the reliability of the evaluation.

\section{Evaluation}
\label{sec:eval}
We begin by visualizing the detection of malicious layers to offer an intuitive understanding of the mechanism (\S\ref{subsec:detect_mal_layers}). Next, we analyze its accuracy in identifying malicious layers (\S\ref{subsec:client_side_check}) and evaluate the system's performance by comparing accuracy, leakage rate, and the number of leaked images with those of other systems. This is followed by an assessment of \DRArmor's defense capabilities compared to other defense mechanisms (\S\ref{subsec:defence}). Additionally, we evaluate the system's performance by comparing metrics such as accuracy, leakage rate, and the number of leaked images with those of other systems. Furthermore, we compute key metrics, including True Positive Rate (TPR), True Negative Rate (TNR), False Positive Rate (FPR), and False Negative Rate (FNR), to assess the mechanism under various experimental settings.

\subsection{Detection of Malicious Layer}
\label{subsec:detect_mal_layers}
 The impact of the placement of malicious layers on the effectiveness of \DRArmor is analyzed below.\par

\begin{figure*}[!t]
    \centering

        \includegraphics[width=0.9\textwidth]{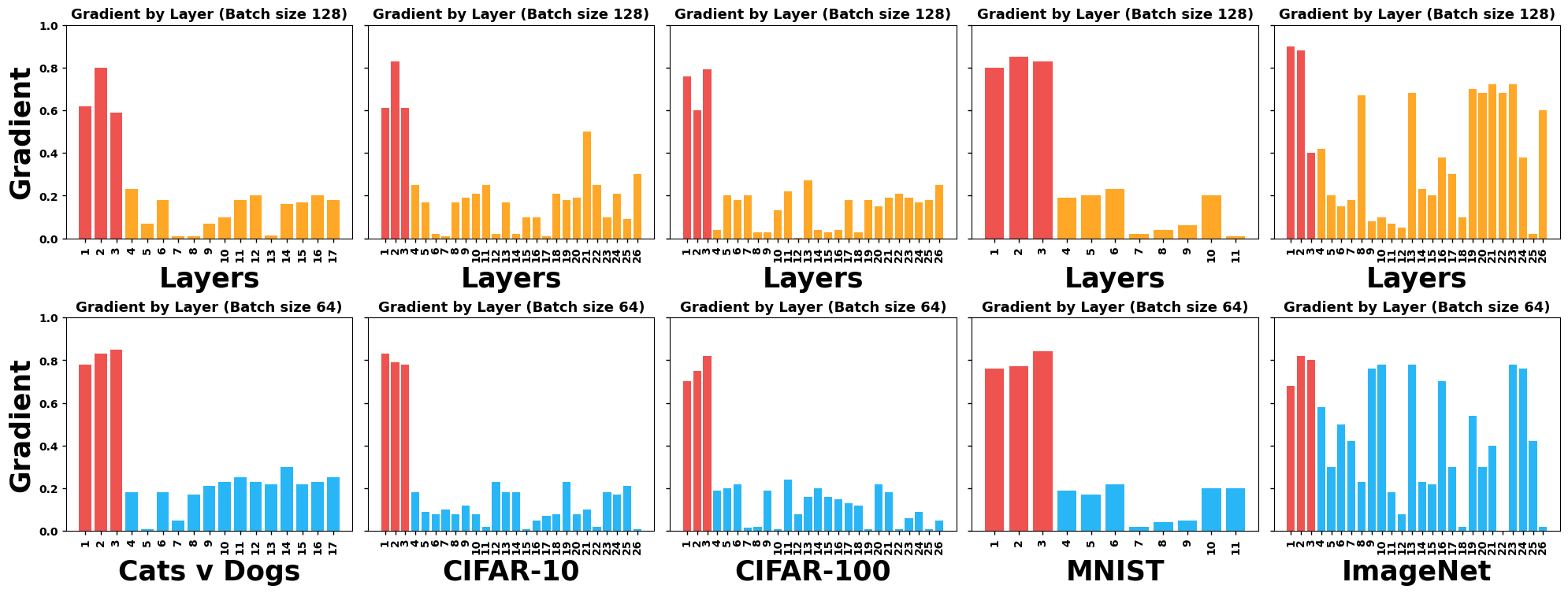}        
    \caption{Gradient Analysis Across Datasets: Identifying Malicious Layers at the Start of the Model using LRP (Batch Sizes: 64, 128) with threshold value of 0.5.}
    \label{fig:bar-graph-lrp}
\end{figure*}
\textbf{Malicious Layers at the Start of the Model.} The original model architecture was modified to include convolutional layers and FC layers at the beginning. To analyze the behavior of the model, LRP was applied after training the model architecture in the client devices. The training process was carried out using batch sizes of 64 and 128. Experiments were carried out on five datasets -- Cats v Dogs, CIFAR-10, CIFAR-100, ImageNet, and MNIST -- using models described in Section \ref{sec:experiments}. Each architecture contained three malicious layers, as introduced in the LoKI paper. These malicious layers were deliberately designed to learn irrelevant features from the data, thereby disrupting the model’s ability to produce meaningful outputs. The goal of the experiments was to evaluate the effectiveness of LRP in detecting these malicious layers. Upon applying LRP, the method demonstrated its ability to identify malicious layers effectively in most datasets. Figure \ref{fig:bar-graph-lrp} illustrates this result, where the gradient values for the malicious layers were significantly higher compared to those of non-malicious layers. This is because, while the majority of layers contributed meaningfully to the model's output, the malicious layers displayed anomalous behavior by learning features that were irrelevant to the task. The higher scores for these layers indicate their deviation from expected behavior, making them stand out in the relevance analysis.\par
\begin{figure}[htbp]
    \centering

        \includegraphics[width=0.5\textwidth]{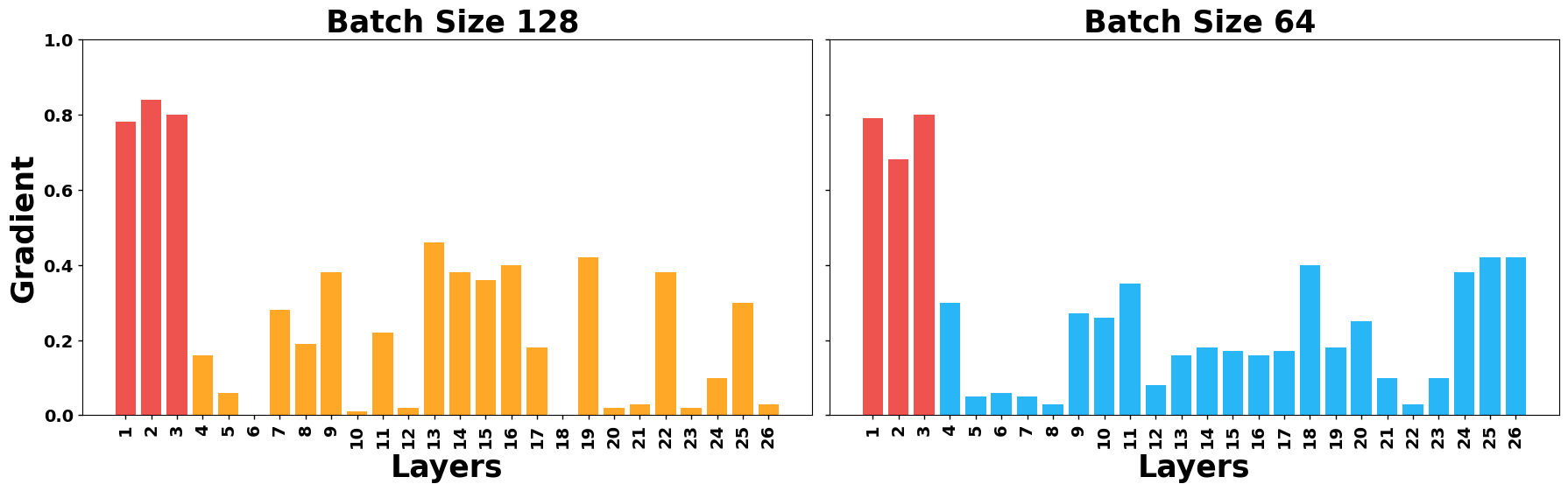}
        
    \caption{Gradient Analysis in ImageNet: Identifying Malicious Layers at the Start of the Model using DTD with threshold value of 0.5.}
    \label{fig:imagenet_dtd}
\end{figure}
The results for the ImageNet dataset were less promising. LRP struggled to effectively differentiate between malicious and non-malicious layers for this dataset. This reduced effectiveness can be attributed to the complexity of the ImageNet dataset, which contains a large number of classes and significantly more diverse and intricate data compared to the other datasets. Models trained on ImageNet require highly detailed feature representations, which can dilute the relevance scores, leading to difficulty in isolating malicious layers. Unlike LRP, DTD provides a more refined decomposition of relevance scores, enabling it to handle complex scenarios more effectively. As shown in Figure \ref{fig:imagenet_dtd}, DTD improved the detection of malicious layers in the ImageNet model. Although the difference between the relevance values of malicious and non-malicious layers was less pronounced compared to simpler datasets, DTD offered a more robust solution for this complex task.\par
The comparison between LRP and DTD highlights the need for adaptive analysis techniques depending on the dataset's complexity. For simpler datasets like Cats v Dogs, CIFAR-10, CIFAR-100, and MNIST, LRP was sufficient to identify malicious layers with high accuracy. In contrast, for the highly complex ImageNet dataset, the limitations of LRP necessitated the use of DTD to achieve better results.\par

\begin{figure*}[!t]
    \centering

        \includegraphics[width=0.9\textwidth]{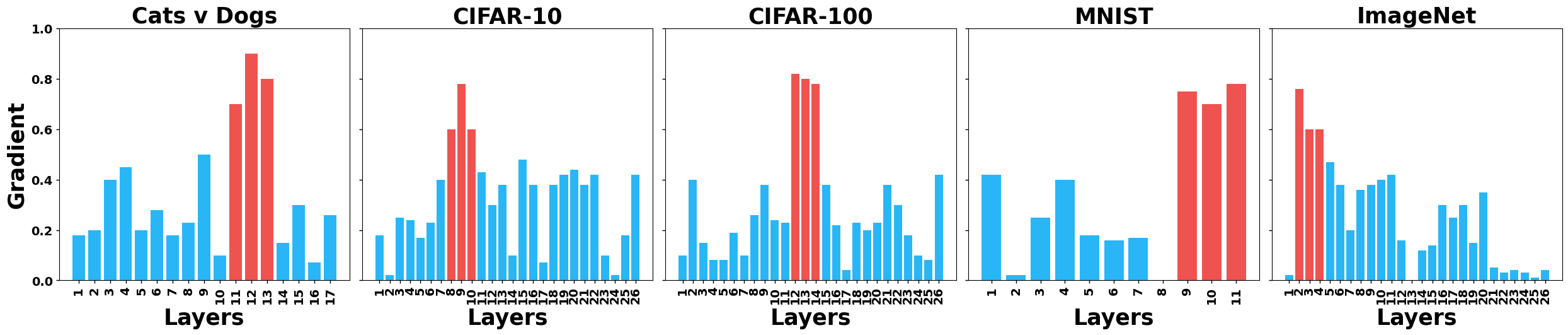}
        
    \caption{Gradient Analysis: Identifying Malicious Layers Further in the Model by \DRArmor (Batch Size 128) with threshold value of 0.5.}
    \label{fig:bar-graph-2}
\end{figure*}

\textbf{Malicious Layers Deeper in the Model.} The placement of malicious layers was randomized and positioned at different depths within the model architecture across various datasets. This design choice was made to emulate real-world scenarios where malicious behavior may not be localized to specific regions of the model but instead distributed unpredictably. The placement strategy ensured that the model's overall accuracy was maintained while introducing the desired anomalies. Figure \ref{fig:bar-graph-2} illustrates the results of this setup, showing how \DRArmor performed in detecting the malicious layers. Unlike the earlier experiments, where malicious layers were consistently positioned at the beginning of the architecture, this randomized placement posed additional challenges for interpretability techniques. Despite these challenges, the methods were able to detect the malicious layers effectively across all datasets, as indicated by the elevated gradients for the malicious layers in Figure \ref{fig:bar-graph-2}. However, compared to the previous setup, the difference in gradient values between malicious and non-malicious layers was notably smaller. This reduced distinction can be attributed to the following factors: 

\begin{itemize}
    \item \textbf{Increased Depth and Dilution of Relevance:} When malicious layers are positioned deeper in the model, the relevance propagation process becomes less distinct due to the accumulation and interaction of gradients from preceding layers. This results in a partial dilution of the relevance scores.
    \item \textbf{Randomized Placement Complexity:} The random placement of malicious layers creates additional variability in the feature maps processed by the model. This variability can mask the anomalous behavior of the malicious layers, reducing the sharpness of their detection.
    \item \textbf{Dependency on Layer Contributions:} As layers further in the model are typically more specialized to work on the learnt features of the previous layers, the ability to isolate gradients depends heavily on how well the interpretability method accounts for these contributions. This can result in smaller differences in gradient values between layers.
\end{itemize}
Despite these reduced distinctions, \DRArmor still performs well in detecting malicious layers. As seen in Figure \ref{fig:bar-graph-2}, the relevance scores for these layers remain consistently higher than those for non-malicious layers. This indicates that, even with increased separation, the techniques retain their capacity to highlight anomalous behaviour within the model.

\subsection{Analyzing Accuracy of Detecting Malicious Layers}
\label{subsec:client_side_check}
Tables \ref{tab:tpr_start_3mal} and \ref{tab:tpr_far_3mal} present the evaluation metrics for the accuracy of the defense mechanism in detecting malicious layers under two scenarios depending on their placement: \textit{the start of the model} and \textit{deeper in the architecture}. The evaluation metrics offer insights into the effectiveness of \DRArmor in correctly identifying malicious and benign layers.\par

\textbf{Performance Across Different Datasets.}
    For all datasets, the TPR and TNR values are relatively high, indicating that the defense mechanism effectively detects most malicious layers and correctly classifies benign ones. MNIST consistently demonstrates the highest TPR and TNR values in both scenarios, with TPR values of 0.980 (start) and 0.906 (middle), and TNR values of 0.989 (start) and 0.941 (middle). This suggests that the simpler nature of MNIST data facilitates more accurate detection of malicious layers.\par
    ImageNet, by contrast, shows lower TPR and TNR values in both cases. For layers at the start, the TPR is 0.746, and the TNR is 0.802. For layers deeper in the model, these drop further to 0.746 and 0.684, respectively. This reduction can be attributed to the increased complexity and diversity of ImageNet, making it more challenging to effectively distinguish malicious from benign layers.\par
\textbf{Effect of Layer Positioning.}
    When malicious layers are positioned at the start of the architecture, the detection mechanism achieves higher TPR and TNR values across most datasets compared to when the layers are deeper in the model. For instance, CIFAR-100 has a TPR of 0.926 and TNR of 0.910 at the start, while these decrease to 0.889 and 0.927, respectively, when the layers are further into the architecture.\par
    This trend suggests that malicious layers located at the start of the architecture are easier to detect. Layers closer to the input are often more general in their feature extraction, making anomalies introduced by malicious layers more apparent. In contrast, layers deeper in the architecture learn more specialized presentations, making it harder to identify malicious activity.\par
\textbf{False Positive and False Negative Rates.}
FPR and FNR remain reassuringly small even with three back‑door layers. As shown in Table \ref{tab:tpr_start_3mal}, at the start of the model it misclassifies $\approx0.11$ benign and misses $0.06$ malicious layers on \textsc{MNIST}, and $4.6$ and $0.76$ layers, respectively, on ImageNet's 26-layer network.  
When the attacker buries the payload deep in the architecture, as shown in Table \ref{tab:tpr_far_3mal}, the worst-case rates on ImageNet rise to $\text{FPR}=0.316$ and $\text{FNR}=0.311$, i.e.\ $\approx7$ false alarms among $23$ benign layers and $<1$ undetected back-door ($0.93$ of 3). All other datasets keep both FPR and FNR below $0.20$, so at least $80\,\%$ of benign layers stay untouched while $\ge74\,\%$ of malicious
insertions are caught. Thus, DRArmor still misses \emph{fewer than one} malicious layer per run on every benchmark, perturbing only a small fraction of the model and maintaining protection. Table ~\ref{tab:impact} confirms that the residual errors have limited consequences: the largest accuracy drop is $4.8$ pp (ImageNet, deep), and the highest SSIM obtained by an attacker is $0.39$, well below the recognisability threshold of $0.45$–$0.50$ as reported in prior inversion work. 
  
To summarise, \DRArmor demonstrates robust performance across all datasets, particularly when malicious layers are located at the start of the model architecture. \DRArmor balances high TPR and TNR values and low FPR and FNR values, ensuring reliable detection with minimal impact on benign layers.

\begin{figure*}[!t]
    \centering
    \begin{subfigure}[b]{\columnwidth}
        \centering
        \includegraphics[width=0.9\textwidth,trim=0cm 0cm 0cm 0.2cm, clip]{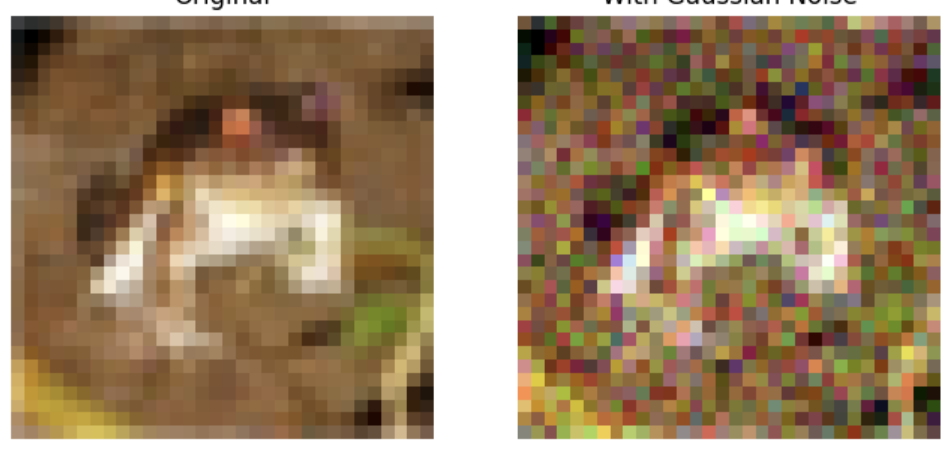}
        \label{fig:image1}
    \end{subfigure}
    \hfill
    \begin{subfigure}[b]{\columnwidth}
        \centering
        \includegraphics[width=0.9\textwidth,trim=0cm 0cm 0cm 0.2cm, clip]{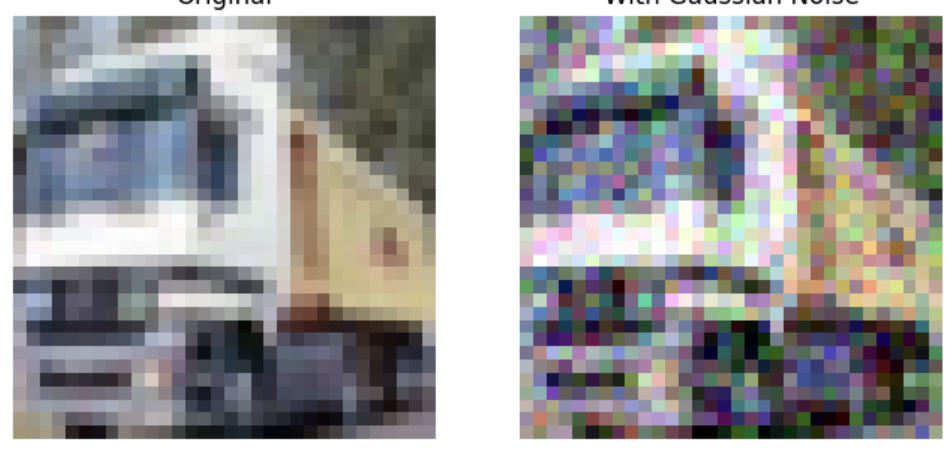}
        \label{fig:image2}
    \end{subfigure}
    \hfill
    \caption{Data Reconstructed at the Server Using DP-Gaussian noise with \( \sigma^2 = 0.2 \) after Identification of the Malicious Layers.}
    \label{fig:all_combined_noise}
    \Description{Data Reconstructed at the Server Using DP-Gaussian with \( \sigma^2 \)=0.2 After Identification of the Malicious Layers.}
\end{figure*}

\subsection{Analysis of \DRArmor Defense}
\label{subsec:defence}
Two strategies may be adopted to mitigate the impact of malicious layers in a model: \textit{adding noise to the gradients} or \textit{completely pruning the malicious layers}. Both approaches aim to protect client data from reconstruction attacks while maintaining the performance of the FL system to the greatest possible extent.\par

\textbf{Impact of Adding Gaussian Noise.} Figure \ref{fig:all_combined_noise} illustrates the impact of applying low Gaussian noise on the gradients. This method perturbs the gradient values, introducing uncertainty that disrupts the ability of malicious layers to perform data reconstruction. Although the added noise significantly reduces the utility of the gradients for machine learning models attempting to interpret them, it does not fully obscure the information from human perception. As shown in the figure, objects in the reconstructed image remain visually identifiable to human observers despite the noise. This highlights a limitation of noise-based methods, as attackers with advanced techniques might still exploit residual patterns to reconstruct sensitive data.\par

\begin{figure*}[!t]
    \centering
    \begin{subfigure}[b]{0.3\textwidth}
        \centering
        \includegraphics[width=\textwidth]{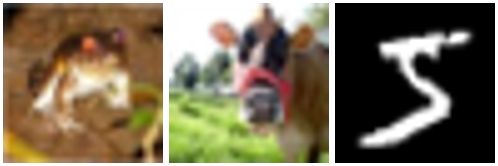}
        \caption{Original Images}
        \label{fig:image1}
    \end{subfigure}
    \hfill
    \begin{subfigure}[b]{0.3\textwidth}
        \centering
        \includegraphics[width=\textwidth]{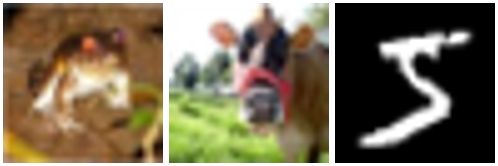}
        \caption{DRA Using LoKI}
        \label{fig:image2}
    \end{subfigure}
    \hfill
    \begin{subfigure}[b]{0.3\textwidth}
        \centering
        \includegraphics[width=\textwidth]{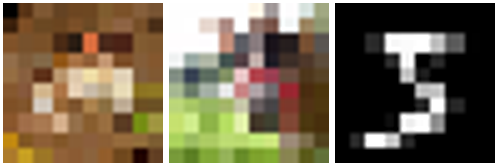}
        \caption{Pixelated Images}
        \label{fig:image2}
    \end{subfigure}
    \hfill
    \caption{Illustration of Reconstruction Results using \DRArmor with Pixelation.}
    \label{fig:pixelated}
    \Description{Illustration of Reconstruction Results Across Multiple Examples. Each Group Contains Three Images: the Original Image (First), the Reconstruction Produced by the DRA Attack Using LoKI (Second), and the Reconstruction Generated when Pixelated Gradients Are Sent from Malicious Layers (Third) After Our Detection Mechanism.}
\end{figure*}
\textbf{Impact of Adding Pixelated Noise.} Figure \ref{fig:pixelated} demonstrates the effect of pixelation on gradients. Pixelation transforms the gradient values into coarse, block-like representations by averaging groups of values within fixed-size blocks. This process effectively removes fine-grained details crucial for reconstruction, replacing them with large, uniform regions. Unlike Gaussian noise, pixelation makes the reconstructed images appear as unrecognizable mosaics of color blocks. While the overall color distribution may be preserved, the structural and spatial details necessary for accurate reconstruction are completely lost. This makes pixelation a stronger defense mechanism against gradient-based DRA, as attackers are unable to retrieve meaningful representations of the original data.\par
\begin{table}[b]
    \caption{FL Task Accuracy Analysis: Comparison of Baseline, \DRArmor with Noise \& Pixelation (NP), and \DRArmor with Pruning at the Start of the Model.}
    \label{tab:acc_start}
    \centering
    \begin{tabular}{|c|c|c|c|}
        \hline
        \textbf{Dataset} & \textbf{FL Acc.} & \textbf{\DRArmor (NP)} & \textbf{\DRArmor (Prune)} \\
        \hline
        MNIST & 0.924 & 0.874 & 0.857 \\
        \hline
        CIFAR-10 & 0.902 & 0.868 & 0.823 \\
        \hline
        CIFAR-100 & 0.894 & 0.843 & 0.796 \\
        \hline
        ImageNet & 0.873 & 0.791 & 0.714 \\
        \hline
        Cats v Dogs & 0.981 & 0.923 & 0.872 \\
        \hline     
    \end{tabular}
\end{table}
\begin{table}[b]
    \caption{FL Task Accuracy Analysis: Comparison of Baseline, \DRArmor with Noise \& Pixelation (NP), and \DRArmor with Pruning Further in the Model.}
    \label{tab:acc_far}
    \centering
    \begin{tabular}{|c|c|c|c|}
        \hline
        \textbf{Dataset} & \textbf{FL Acc.} & \textbf{\DRArmor (NP)} & \textbf{\DRArmor (Prune)} \\
        \hline
        MNIST & 0.924 & 0.824 & 0.813 \\
        \hline
        CIFAR-10 & 0.902 & 0.829 & 0.789 \\
        \hline
        CIFAR-100 & 0.894 & 0.794 & 0.755 \\
        \hline
        ImageNet & 0.873 & 0.767 & 0.708 \\
        \hline
        Cats v Dogs & 0.981 & 0.870 & 0.853 \\
        \hline     
    \end{tabular}
\end{table}
\textbf{Impact of Pruning.} The impact of pruning malicious layers is summarized in Tables \ref{tab:acc_start} and \ref{tab:acc_far}. Pruning removes entirely the identified malicious layers, eliminating their ability to contribute to DRAs. This approach balances accuracy and privacy, achieving 86\% accuracy for MNIST and 80\% for CIFAR-100, even with enhanced privacy measures. While there is a slight performance trade-off compared to noise addition, it leaves no space for data leakage as discussed in \S\ref{subsec:leakage_rate}. \par

\textbf{Impact of False Positives and False Negatives on Defense.}  Tables \ref{tab:tpr_start_3mal}–\ref{tab:impact} show that DRArmor’s false alarms (FP) and misses (FN) have tightly bounded consequences.
A missed back-door reveals at most the gradients of that single layer. At the same time, a spurious alarm adds controlled noise or prunes a benign layer, producing a recoverable dip in accuracy.  The server
handles both cases gracefully: updates with the correct shape are accepted, and any malformed update is quietly discarded, preventing knock-on effects. As the tables indicate, occasional misclassifications introduce some information exposure and accuracy loss, but these effects stay within the modest bounds detailed in the
tables and do not materially compromise overall privacy or utility.
\begin{table}[!b]
\caption{Detection reliability when \textbf{three} malicious layers are inserted at the \textbf{start} of the model.}
\label{tab:tpr_start_3mal}
\centering
\resizebox{\linewidth}{!}{%
\begin{tabular}{|c|c|c|c|c|c|c|c|c|}
\hline
\textbf{Dataset} & \textbf{$B$} & \textbf{$M$} & \textbf{TPR} & \textbf{TNR} & \textbf{FPR} & \textbf{FNR} & \textbf{\#FP} & \textbf{\#FN}\\
\hline
MNIST        &  8 & 3 & 0.980 & 0.989 & 0.011 & 0.020 & 0.09 & 0.06\\ \hline
CIFAR‑10     & 23 & 3 & 0.913 & 0.924 & 0.076 & 0.087 & 1.75 & 0.26\\ \hline
CIFAR‑100    & 23 & 3 & 0.926 & 0.910 & 0.090 & 0.074 & 2.07 & 0.22\\ \hline
ImageNet     & 23 & 3 & 0.746 & 0.802 & 0.198 & 0.254 & 4.55 & 0.76\\ \hline
Cats v Dogs & 14 & 3 & 0.842 & 0.859 & 0.141 & 0.158 & 1.97 & 0.47\\
\hline
\end{tabular}}
\end{table}

\begin{table}[!b]
\caption{Detection reliability when \textbf{three} malicious layers are inserted \textbf{deep} in the model.}
\label{tab:tpr_far_3mal}
\centering
\resizebox{\linewidth}{!}{%
\begin{tabular}{|c|c|c|c|c|c|c|c|c|}
\hline
\textbf{Dataset} & \textbf{$B$} & \textbf{$M$} & \textbf{TPR} & \textbf{TNR} & \textbf{FPR} & \textbf{FNR} & \textbf{\#FP} & \textbf{\#FN}\\
\hline
MNIST        &  8 & 3 & 0.906 & 0.941 & 0.059 & 0.094 & 0.47 & 0.28\\ \hline
CIFAR‑10     & 23 & 3 & 0.894 & 0.910 & 0.090 & 0.106 & 2.07 & 0.32\\ \hline
CIFAR‑100    & 23 & 3 & 0.889 & 0.927 & 0.073 & 0.111 & 1.68 & 0.33\\ \hline
ImageNet     & 23 & 3 & 0.746 & 0.684 & 0.316 & 0.311 & 7.27 & 0.93\\ \hline
Cats v Dogs & 14 & 3 & 0.800 & 0.814 & 0.186 & 0.200 & 2.60 & 0.60\\ 
\hline
\end{tabular}}
\end{table}

\begin{table}[!b]
\caption{Effect of misclassifications on utility and privacy when \textbf{three} malicious layers are injected per run.
Utility $\Delta$ = drop in validation accuracy caused by false-positive (FP) perturbations.
Privacy $\Delta$ = SSIM of reconstructions enabled by a false negative (FN); higher means more leakage.}
\label{tab:impact}
\centering
\resizebox{\linewidth}{!}{%
\begin{tabular}{|c|c|c|c|c|}
\hline
\textbf{Dataset} & \multicolumn{2}{c|}{\textbf{Start}} & \multicolumn{2}{c|}{\textbf{Deep}}\\
\cline{2-5}
& Utility $\Delta$ (\%) & Privacy $\Delta$ (SSIM) & Utility $\Delta$ (\%) & Privacy $\Delta$ (SSIM)\\
\hline
MNIST        & \(-0.3\) & 0.09 & \(-0.6\) & 0.13\\ \hline
CIFAR-10     & \(-0.8\) & 0.14 & \(-1.4\) & 0.20\\ \hline
CIFAR-100    & \(-1.1\) & 0.16 & \(-2.0\) & 0.24\\ \hline
ImageNet     & \(-2.6\) & 0.27 & \(-4.8\) & 0.39\\ \hline
Cats v Dogs  & \(-1.4\) & 0.20 & \(-2.6\) & 0.26\\
\hline
\end{tabular}}
\end{table}

\subsection{Impact of \DRArmor on FL Accuracy}
Eliminating malicious layers (\S\ref{subsec:defence}) is designed to have minimal impact on the model's overall accuracy, as these layers contribute little to the overall output of the model. The aim is to evaluate how defense mechanisms, such as adding noise or pruning, impact the model’s accuracy. The result for accuracy is shown in Tables \ref{tab:acc_start} and \ref{tab:acc_far} for the scenarios, respectively. The first column displays the baseline accuracy of the model without any malicious layers. This represents the optimal accuracy achieved when the model is fully functional and unaffected by adversarial layers. The second column shows the model's accuracy after applying the defense mechanism by adding noise to the gradients sent by the client. The third column reflects the accuracy after pruning the malicious layers. It is observed that the model can predict the output with an accuracy of 79 to 92\% for the data sets used in the experiment. \par
\begin{figure}[!t]
    \centering
    \begin{subfigure}[b]{0.4\columnwidth}
        \centering
        \includegraphics[width=\textwidth]{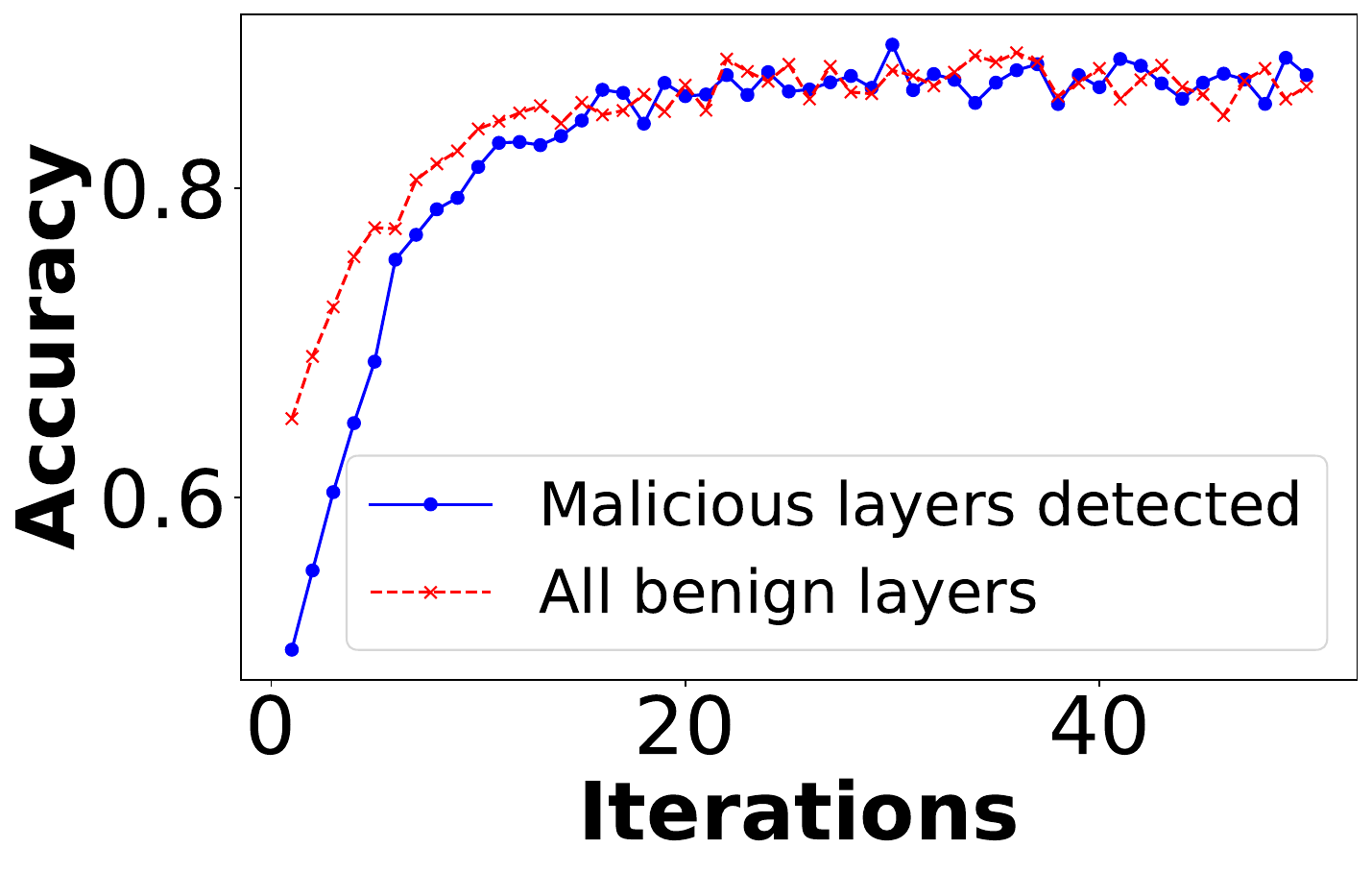}
        \caption{Accuracy}
        \label{fig:image1}
    \end{subfigure}
    \hfill
    \begin{subfigure}[b]{0.4\columnwidth}
        \centering
        \includegraphics[width=\textwidth]{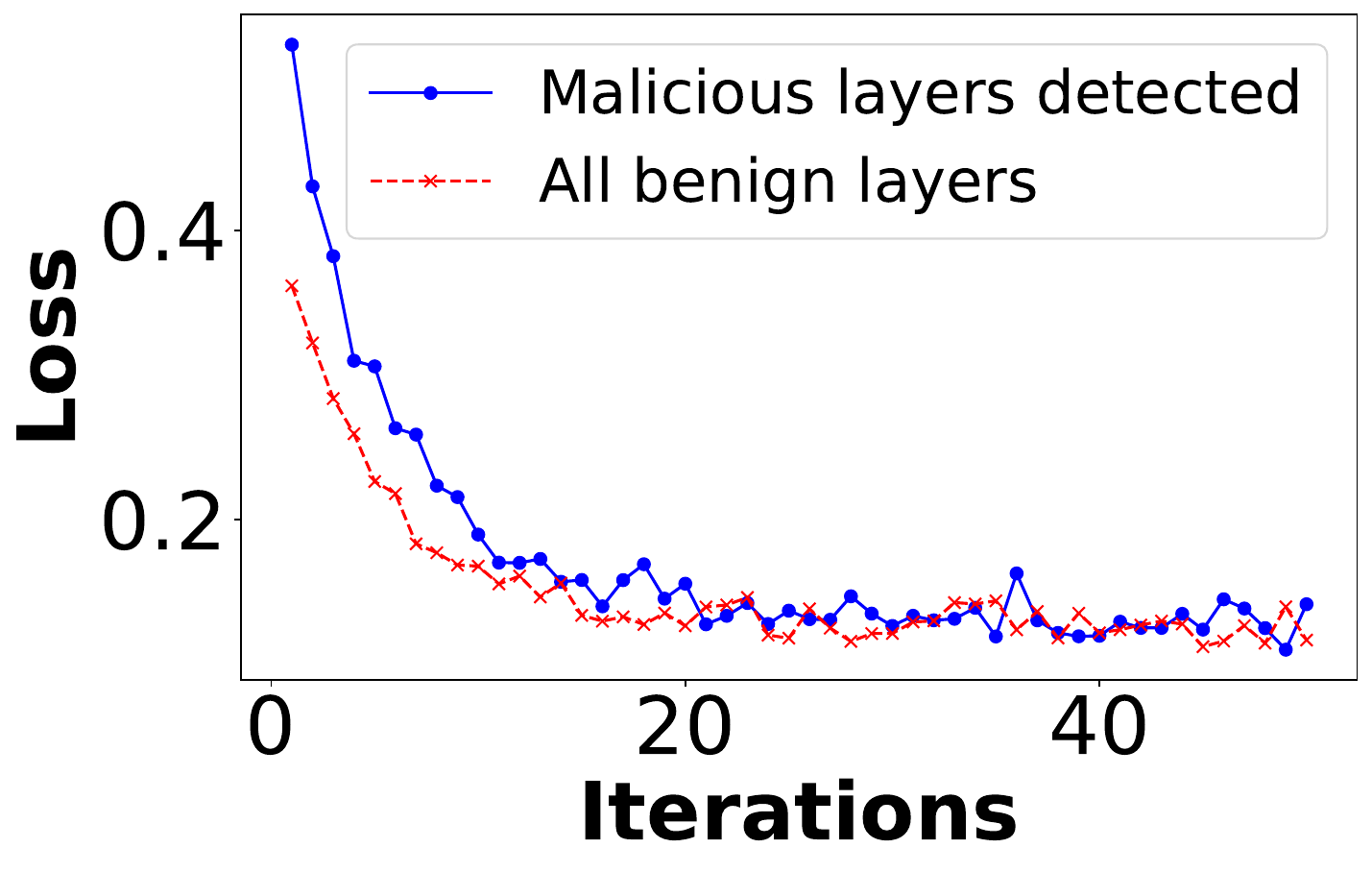}
        \caption{Loss}
        \label{fig:image2}
    \end{subfigure}
    \hfill
    \caption{FL Task Accuracy and Loss Comparison of the MNIST Aggregated Model: Without Malicious Layers vs. With Detected and Mitigated Malicious Layers.}
    \label{fig:acc_loss}
    \Description{FL Task Accuracy and Loss Comparison of the Aggregated Model of MNIST Under Two Scenarios: Without Malicious Layers and with Malicious Layers Detected and Mitigated.}
\end{figure}

\begin{figure}[!t]
    \centering
    \includegraphics[width=0.4\columnwidth]{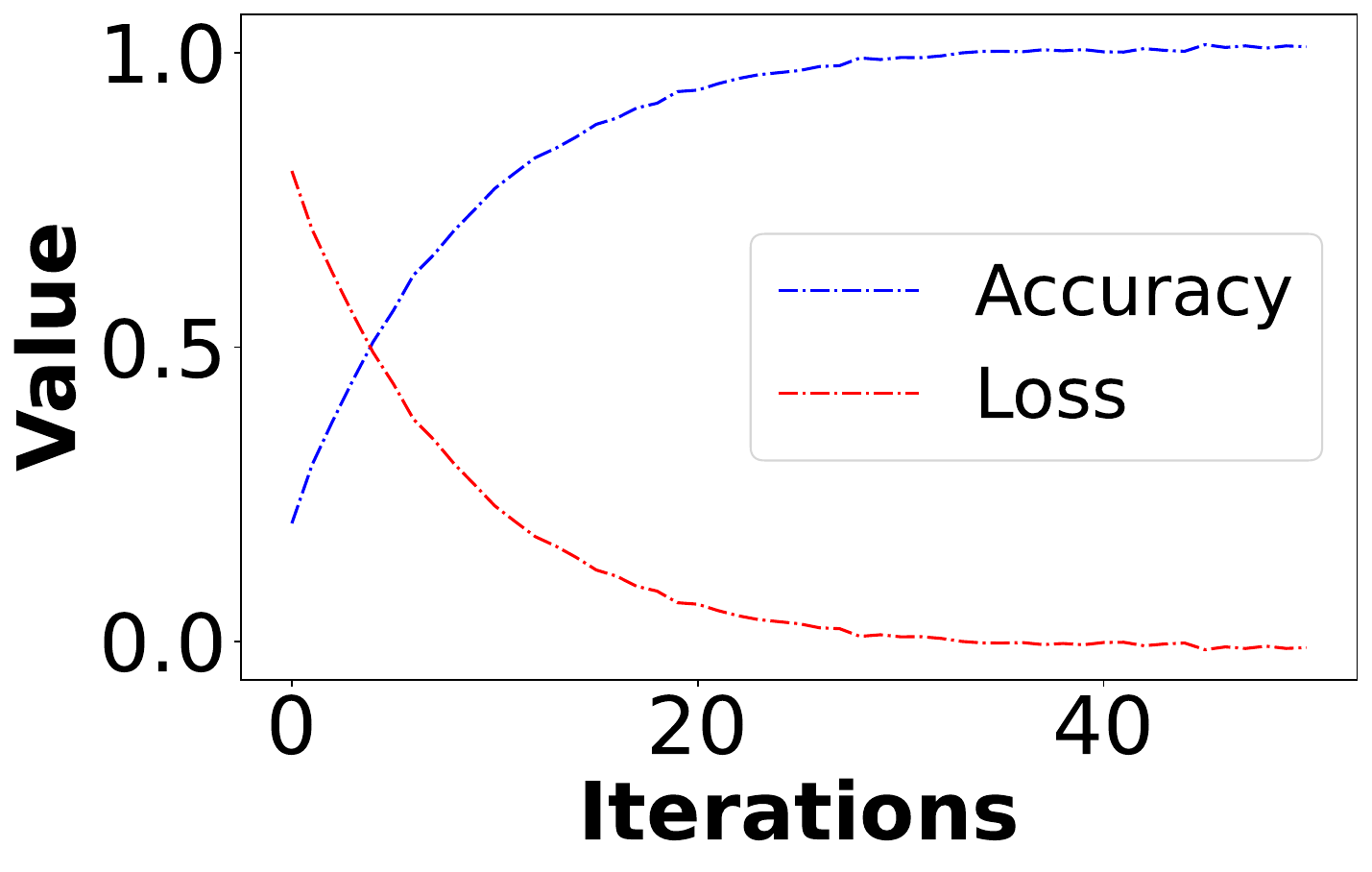}
    \caption{FL Task Accuracy and Loss Comparison of the Aggregated Model on MNIST.}
    \label{fig:far_acc_loss}
    \Description{FL Task Accuracy and Loss Comparison of the MNIST Aggregated Model: Without Malicious Layers vs. With Random Malicious Layers Detected and Mitigated.}
\end{figure}
Moreover, Figures \ref{fig:acc_loss} and \ref{fig:far_acc_loss} illustrate the progression of accuracy and loss during training under two conditions: (1) when the model contains only benign layers, and (2) when malicious layers are detected, and defense mechanisms are applied. The initial trajectories of accuracy and loss differ between the two cases of layer positioning, reflecting the disruption caused by malicious layers. However, as training progresses and the defense mechanisms are applied, the metrics gradually converge. For the MNIST dataset, this common convergence point reaches approximately 87\%, demonstrating that the defense mechanisms successfully restore the model's performance to a level comparable to that achieved without malicious layers.\par
While the defense mechanisms improve accuracy, they are not perfect due to the non-zero chance of false positives in the detection mechanism. Occasionally, the system may mistakenly identify a non-malicious layer as malicious. When this occurs, either noise is added to, or pruning is applied to a legitimate layer, which can impact the model's performance. As seen in Tables  \ref{tab:tpr_start_3mal} and \ref{tab:tpr_far_3mal}, these cases are infrequent and contribute only marginally to accuracy reduction.
\begin{figure}[!t]
    \centering
    \begin{subfigure}[b]{0.49\columnwidth}
        \centering
        \includegraphics[width=\textwidth]{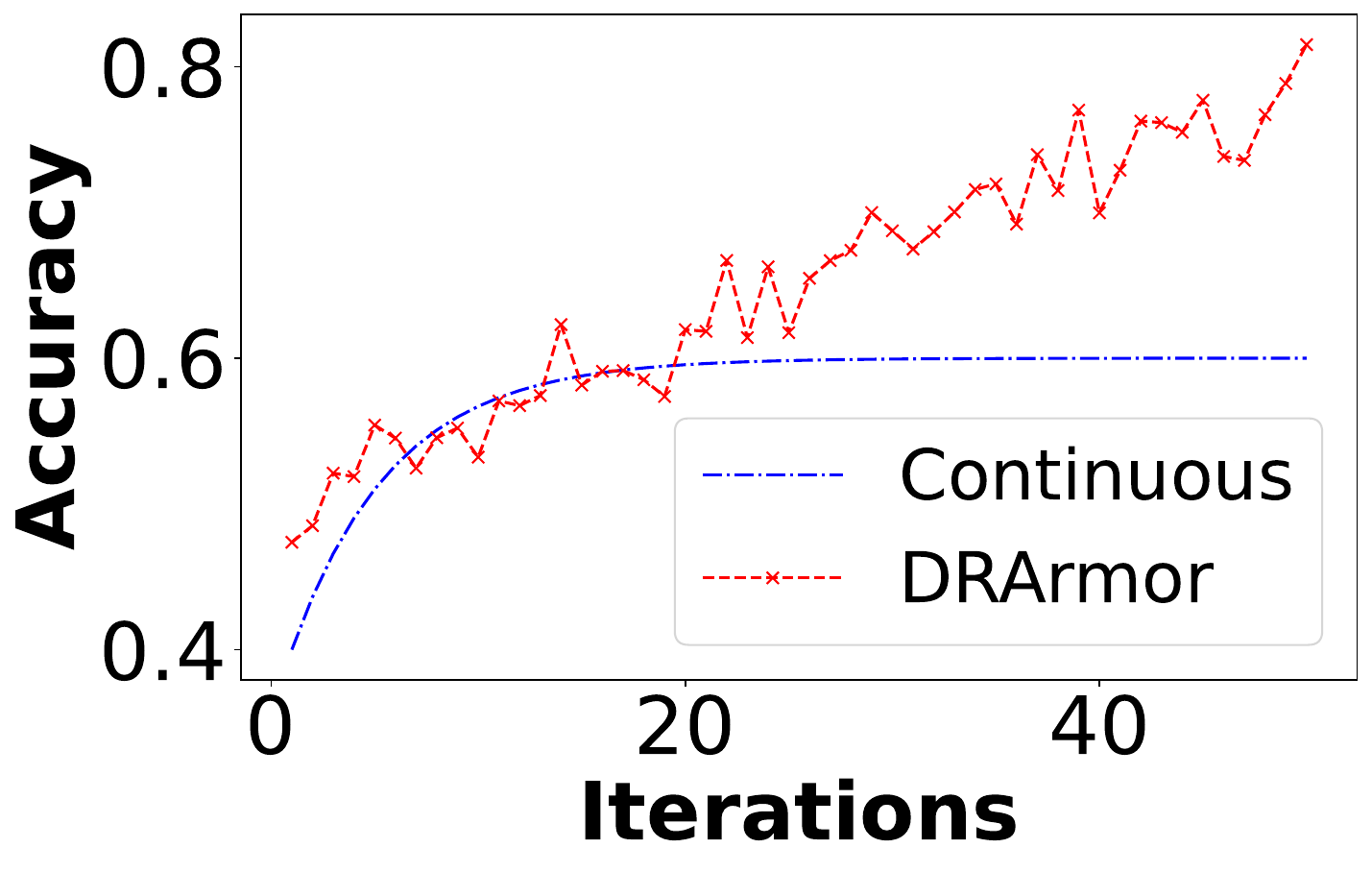}
        \caption{Continuous Poisoning}
        \label{fig:image1}
    \end{subfigure}
    \hfill
    \begin{subfigure}[b]{0.49\columnwidth}
        \centering
        \includegraphics[width=\textwidth]{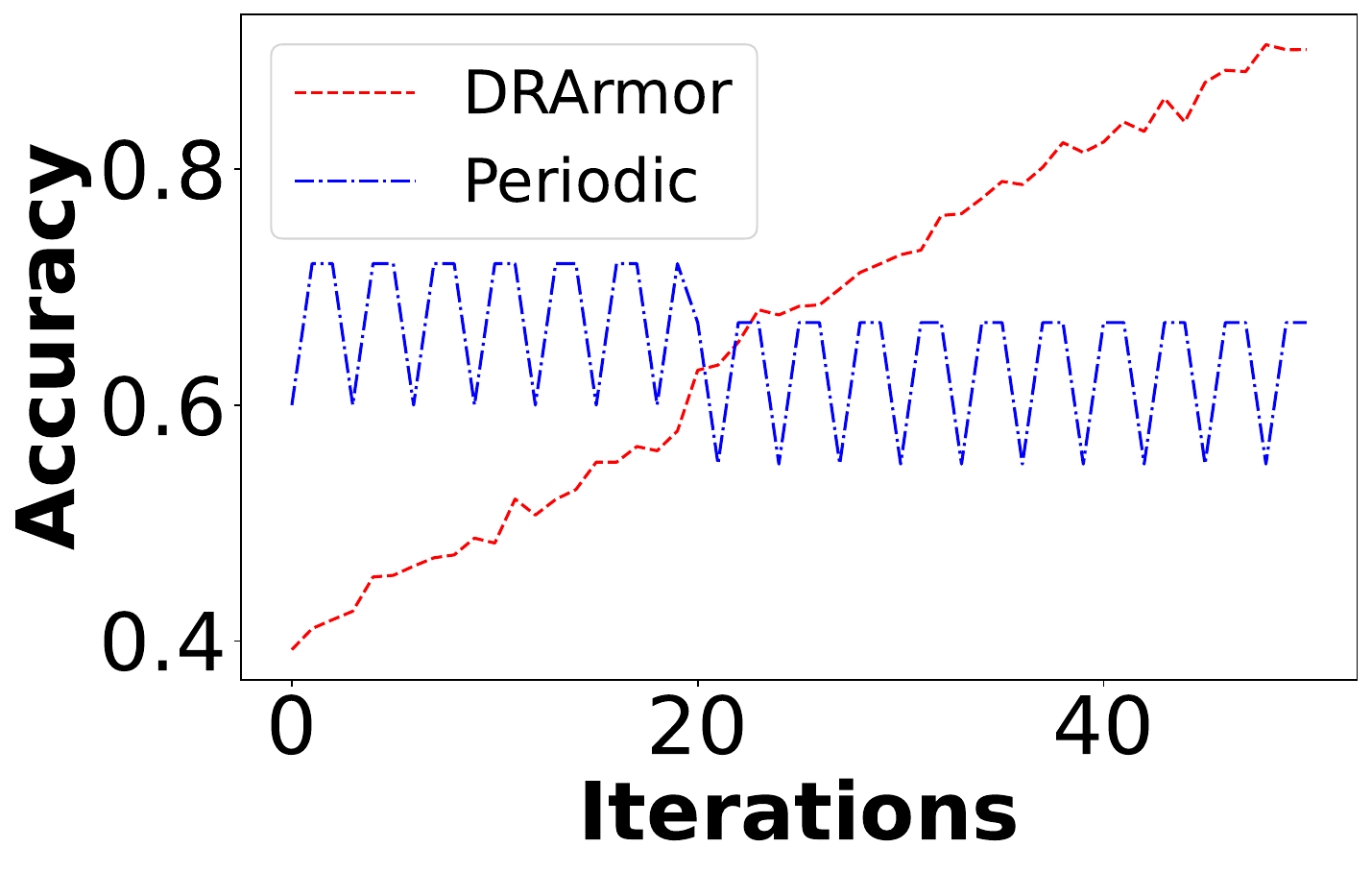}
        \caption{Periodic Poisoning}
        \label{fig:image2}
    \end{subfigure}
    \caption{Comparison of Poisoning Impact on Accuracy for Continuous and Periodic Poisoning vs. Malicious Layers Mitigated.}
    \label{fig:poisoning}
    \Description{Comparison of poisoning impact on accuracy for continuous and periodic poisoning vs. malicious layers removed}
\end{figure}
\subsection{Impact of Poisoning on Accuracy}
\label{subsec:poisoning}
As an extension of the experimentation, we examined the impact of continuous and periodic poisoning on the accuracy of \DRArmor. This metric is not available in related solutions such as \cite{sun2021soteria} and \cite{fan2024guardian}. \par 
In an FL setting, an adversary can inject malicious layers into the model either continuously throughout the training process or during specific, random rounds of training. These scenarios mimic real-world attack strategies where malicious behavior may be persistent or sporadic to evade detection. To assess the robustness of our approach, the client nodes were evaluated for their ability to detect malicious layers, regardless of whether the injection occurred continuously or periodically. The goal was to determine whether the type and timing of poisoning affected the model's ability to identify and mitigate malicious activity. Figure \ref{fig:poisoning} illustrates the relationship between the poisoning strategy and the accuracy of \DRArmor. The graph contains two lines: the poisoning line, representing the frequency and pattern of malicious layer injections (continuous or periodic); and the accuracy line, showing the model's accuracy over time during training. The figure shows that the accuracy of \DRArmor remains unaffected by the type of poisoning strategy. With the proposed detection and mitigation mechanism, the accuracy begins to increase until it stabilises steadily at a consistent level. 

\subsection{Impact of \DRArmor on Leakage Rate}
\label{subsec:leakage_rate}
\begin{figure}[!t]
    \centering
    \begin{subfigure}[b]{0.495\columnwidth}
        \centering
        \includegraphics[width=\textwidth]{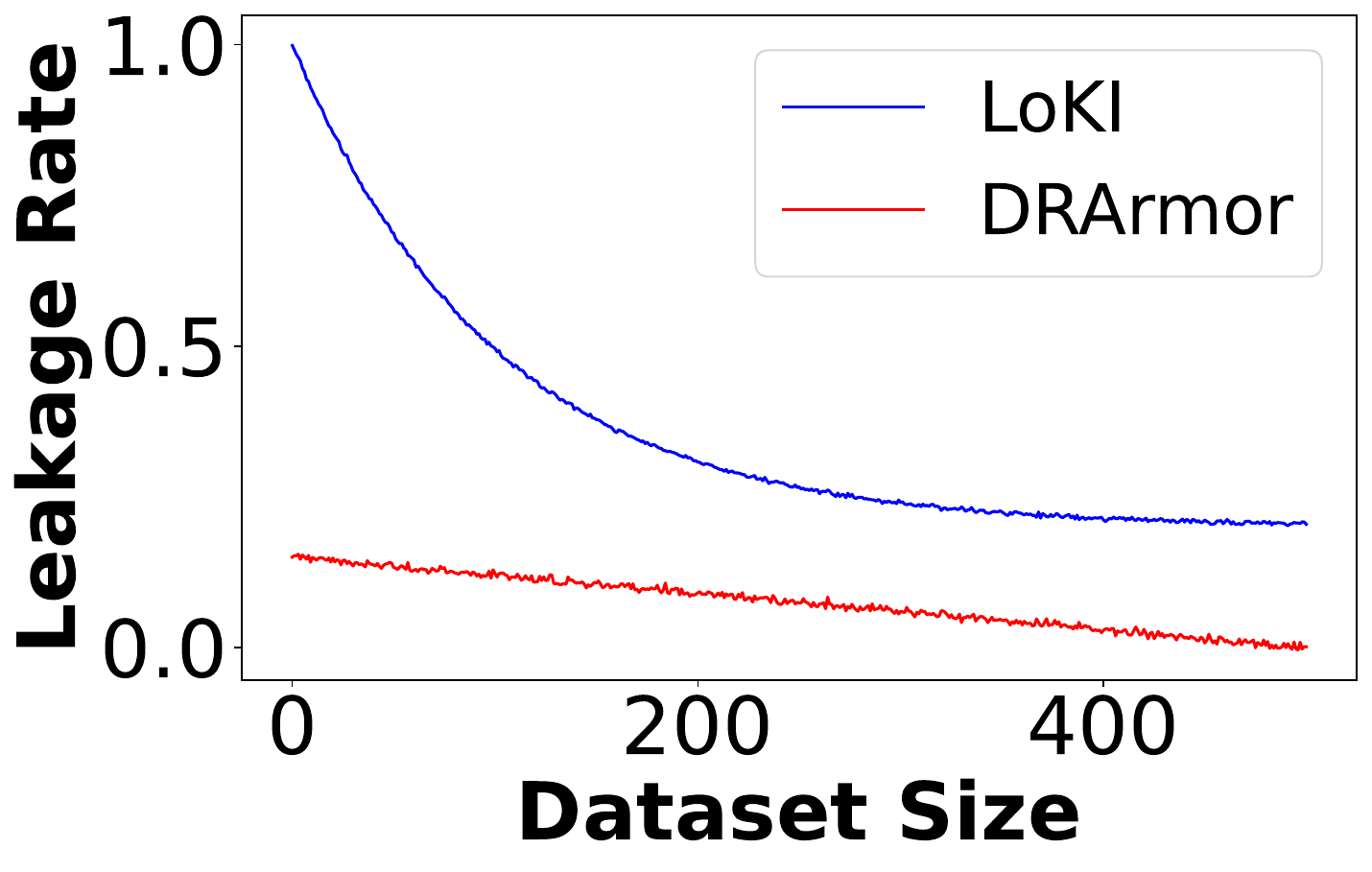}
        \caption{Leakage Rate}
        \label{fig:leakage_rate}
    \end{subfigure}
    \hfill
    \begin{subfigure}[b]{0.495\columnwidth}
        \centering
        \includegraphics[width=\textwidth]{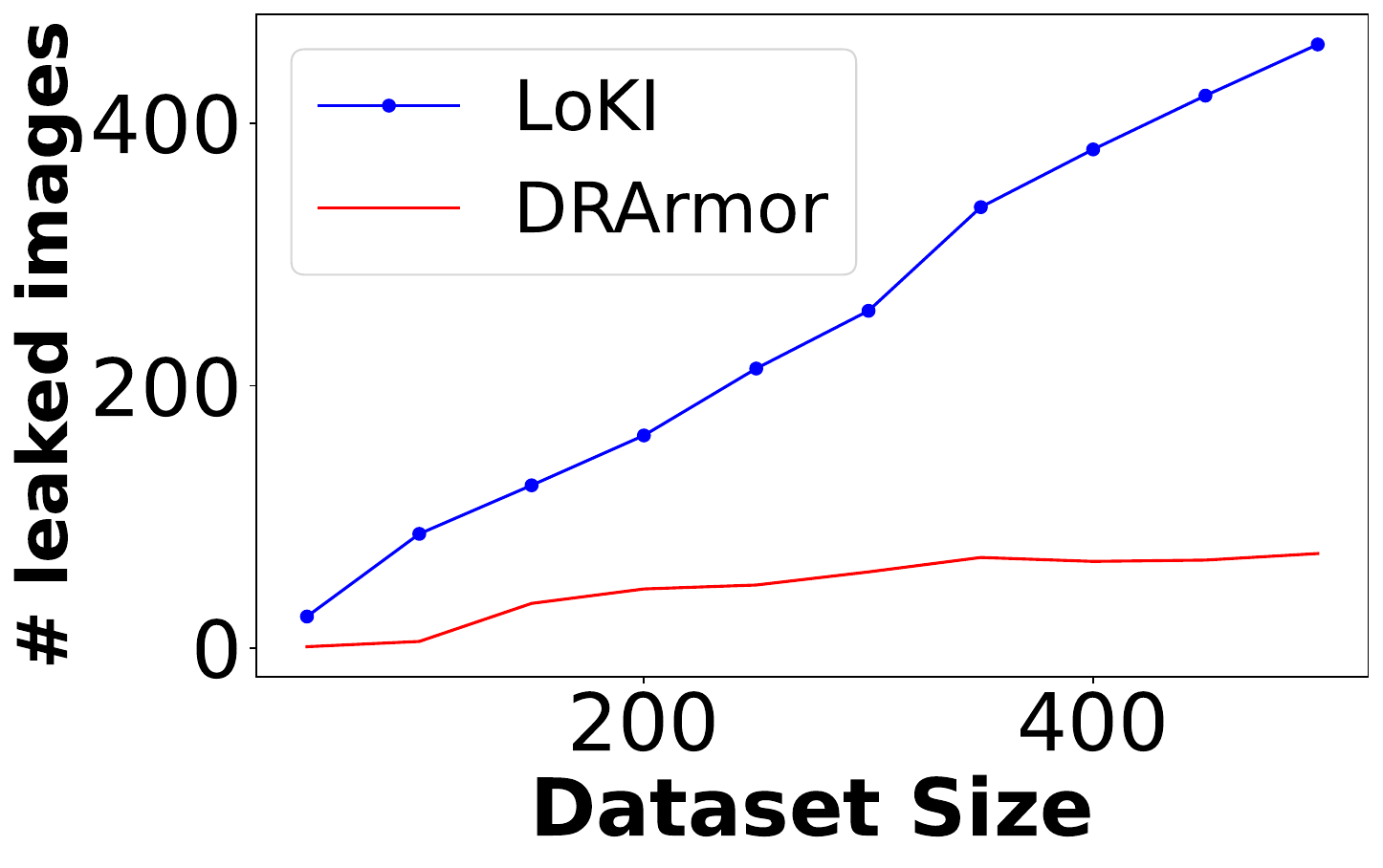}
        \caption{Number of Images Leaked}
        \label{fig:leaked_images}
    \end{subfigure}
    \caption{Leakage Rate and Number of Leaked Images as a Function of the Local Dataset Size Averaged over 200 Clients.}
    \label{fig:leak}
    \Description{Leakage Rate and Number of Leaked Images as a Function of the Local Dataset Size averaged over 200 Clients.}
\end{figure}

The leakage rate is primarily influenced by two factors: \textit{the size of the local dataset} and \textit{the size of the FC layers}. Our experiment aimed to identify malicious layers and mitigate leakage from these layers using two approaches: noise addition and pruning. Figure \ref{fig:leak} shows the effect of the dataset size on the leakage rate. \par
When the malicious layers are pruned, the system ensures that no gradients from these layers contribute to the updates. This effectively reduces the leakage rate to 0\%, as no information from the malicious layers is propagated. When noise is added to the gradients of malicious layers, the leakage rate can vary depending on the magnitude of the noise applied. A higher noise level generally results in a lower leakage rate, but some information may still be leaked if the noise is insufficient. The system's effectiveness in detecting malicious layers directly influences the leakage rate. The leakage rate is significantly reduced if the malicious layers are correctly identified. If detection is inaccurate (false negatives), leakage may occur, as malicious layers remain unmitigated. With an accuracy range of 72\%--87\%, some leakage is observed, particularly as the local dataset size increases. However, this leakage is minimal compared to the baseline leakage rate observed in the LoKI framework. The experiment confirms that \DRArmor reduces the leakage rate to a minimal level. \par

\begin{figure}[!t]
    \centering
 
        \includegraphics[width=\columnwidth]{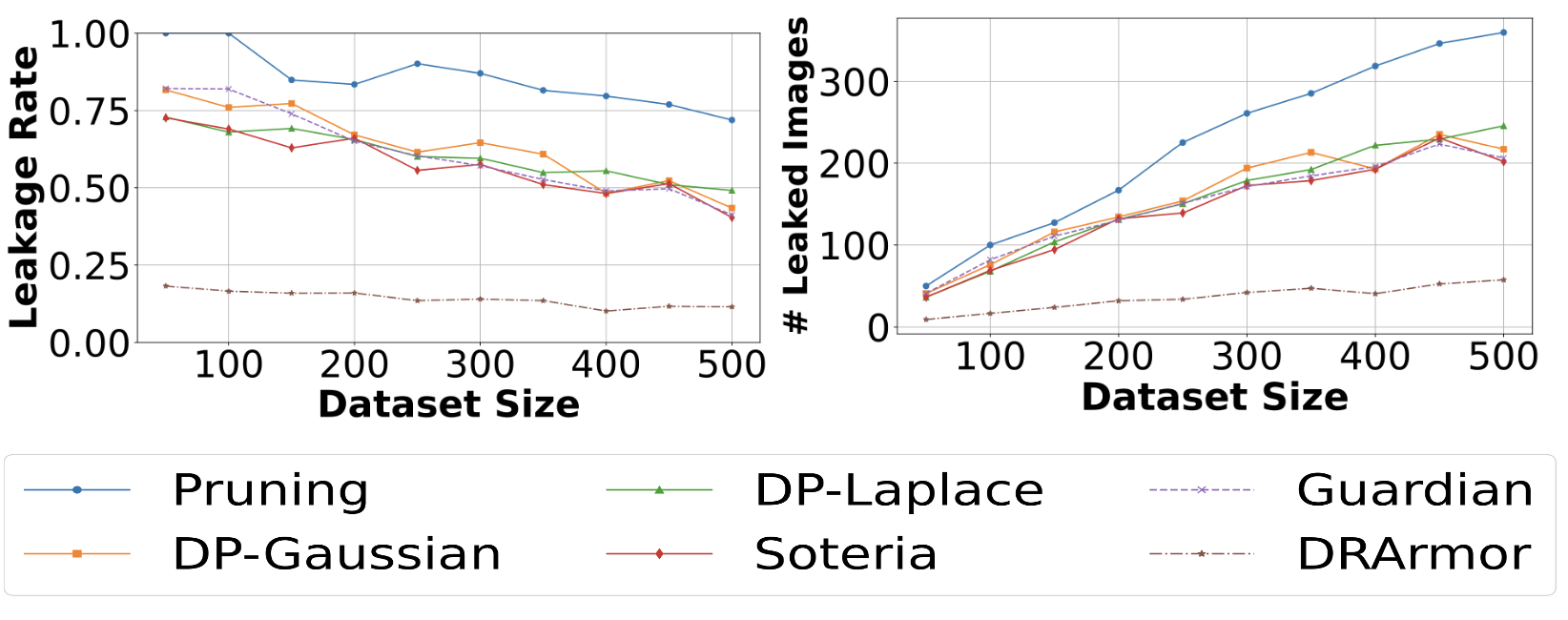}
    
    \caption{Leakage Rate and Number of Leaked Images as a Function of the Local Dataset Size Averaged over 200 Clients against Other Defenses.}
    \label{fig:comparison}
    \Description{Leakage Rate and Number of Leaked Images as a Function of the Local Dataset Size Averaged over 200 Clients for All the Other Defense Methods.}
\end{figure}

\begin{table}[ht]

\centering
\caption{Comparison of DRArmor with existing defenses on CIFAR-100}
\label{tab:defense-results}
\begin{tabular}{|c|c|c|}
\hline
\textbf{Defense Method} & \textbf{Privacy} & \textbf{Utility} \\ \hline
\textbf{No Defense} & 74.8\% / 312 ± 7.4 & 92.5 ± 0.4 \\ \hline
\textbf{DP-Gaussian} & 19.5\% / 68 ± 3.4 & 82.1 ± 0.6 \\ \hline
\textbf{DP-Laplace} & 26.4\% / 92 ± 4.1 & 80.3 ± 0.7 \\ \hline
\textbf{Soteria} & 41.8\% / 154 ± 6.0 & 82.4 ± 0.5 \\ \hline

\textbf{DRArmor} & \textbf{12.2\% / 44 ± 2.3} & \textbf{86.7 ± 0.5} \\ \hline
\end{tabular}
\end{table}

\section{Discussion}
\label{sec:discussion}

\textbf{Comparison with Other Defense Algorithms}
Previous research (\S\ref{sec:relwork}) focuses on mitigating the effects of gradients sent by malicious models to the server, which the DRA exploits. Although these methods attempt to obscure or alter the gradients to prevent reconstruction, they fail to address the root cause of these attacks. \DRArmor takes a fundamentally different direction by identifying the malicious layers and model parameters responsible for leaking gradients to reconstruct the client device data. \DRArmor prevents gradient leakage and ensures that malicious layers causing leakage are effectively detected and eliminated (as shown in \S\ref{subsec:defence}). \par

The new approach in \DRArmor makes directly comparing with related work challenging. Inspired by \cite{zhao2024LoKI}, to evaluate its effectiveness, we compare its performance against existing defences based on leakage rate and the number of images leaked per dataset size. 
As shown in Figure \ref{fig:comparison}, \DRArmor achieves a significantly lower leakage rate compared to other methods replicated in our evaluations (e.g. \cite{sun2021soteria,fan2024guardian}). 
 
Moreover, table~\ref{tab:defense-results} compares DRArmor with baseline defenses in terms of privacy and utility. Privacy is quantified using the leakage rate and number of reconstructed images, while utility is measured by test accuracy while defending against the DRA. DRArmor achieves the lowest privacy leakage with minimal accuracy degradation, outperforming methods like DP-SGD and Soteria, which apply global obfuscation without identifying the malicious layers responsible for leakage.
 \par

Another notable strength of our approach is its robustness to changes in attack architecture. Traditional defense mechanisms are often compromised when the attack model changes, but our method remains effective due to its use of Explainable AI. As shown in Figure \ref{fig:bar-graph-2}, \DRArmor can identify malicious layers even when the model architecture changes, providing flexibility and resilience not feasible in related work. \par

The threat model used in \DRArmor is based on a real-world scenario where an attacker can inject malicious models at any iteration, where the client must detect the attack as soon as it occurs. We evaluated \DRArmor with continuous and periodic poisoning attacks to test this, achieving consistent accuracy in both cases \S\ref{subsec:poisoning}. We also assessed the scalability of \DRArmor by testing it across multiple clients. Unlike the studies in the attacks\cite{zhao2024LoKI,fowl2021robbing}, which used 100 clients, our experiments scale up to 200 clients to evaluate the impact of these attacks on model accuracy in a larger setting.\par

\textbf{Limitations of \DRArmor}:
While our method is the first to detect DRAs in FL, it currently relies on a limited set of XAI techniques for identifying malicious layers. Although we demonstrate the effectiveness of DRArmor across both small and large models, distinguishing between malicious and non-malicious layers can become less precise as model complexity increases. To improve robustness and interpretability, more advanced or domain-adapted attribution methods (e.g.,\cite{selvaraju2016grad,ribeiro2016should}) may offer better explanatory power. DRArmor is designed to be modular, so such techniques can be substituted into our framework depending on deployment needs. A discussion on the computational practicality of DRArmor’s components, including their applicability to typical federated clients, is provided in \S\ref{sec:methodology} and Appendix \ref{subsec:scalability}. We leave the exploration of alternative XAI methods and extended scalability analysis as directions for future work.

Lastly, while \DRArmor's detection mechanism reduces data leakage, it also leads to a slight decrease in the model's accuracy compared to the original model. However, considering our primary objective is preserving privacy in FL systems, we argue this trade-off is acceptable given significantly improved privacy. \par

\section{Conclusion}
\label{sec:conclusion}
We present \DRArmor, a robust defense mechanism designed to detect and mitigate Data Reconstruction Attacks in FL. To the best of our knowledge, \DRArmor is the first solution that leverages Explainable AI to effectively analyze how individual layers of a model contribute to its output. By identifying layers that are not learning relevant features but still send disproportionately high gradients, \DRArmor can accurately classify these layers as malicious. Once detected, \DRArmor employs techniques such as noise injection, pixelation, and pruning to neutralize the impact of these malicious layers, safeguarding client privacy without compromising model integrity. Unlike the studies in \cite{zhao2024LoKI} and \cite{fowl2021robbing}, which used 100 clients, we replicated these attacks with 200 clients to evaluate the effectiveness of \DRArmor in larger settings. Our results demonstrate that \DRArmor significantly reduces data leakage, even as the number of client nodes and the size of the dataset increase. Compared to existing defense algorithms, \DRArmor achieves an average accuracy rate of 87\%, confirming the efficacy of our proposed solution. 
\newpage

\bibliographystyle{ACM-Reference-Format}
\bibliography{bibtex}

\appendix
\begin{figure*}[t!] 
    \centering
    \includegraphics[width=\textwidth]{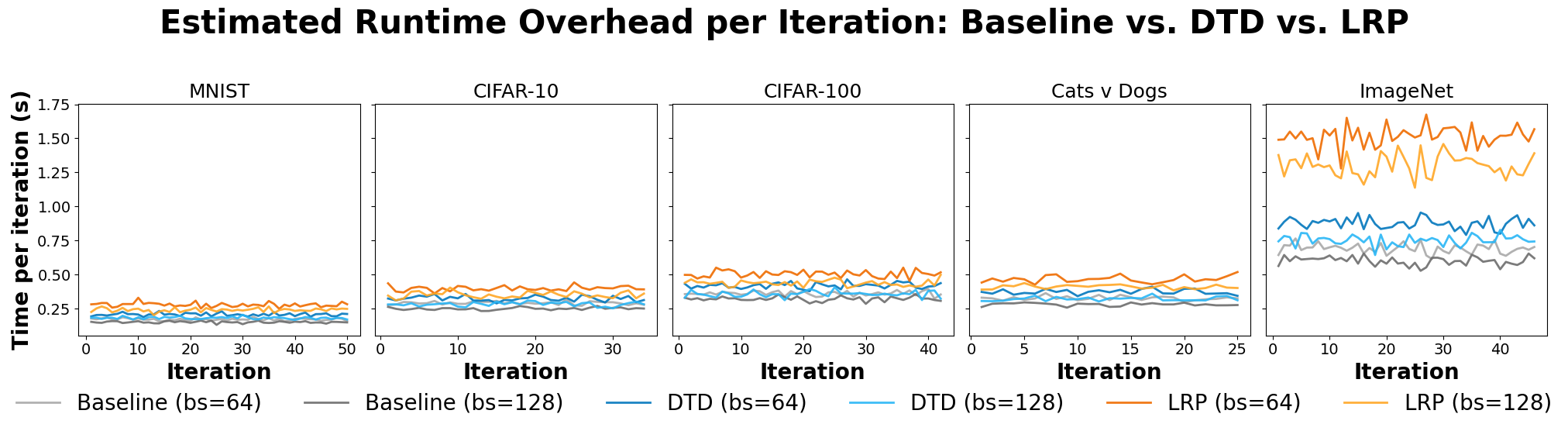}
    \caption{Estimated per-iteration runtime overhead: baseline training vs.DRArmor with DTD and LRP, at batch sizes 64 and
128, across models of varying complexity.}
    \label{fig:runtime-overhead}
    \Description{}
\end{figure*}

\begin{figure*}[b!] 
    \centering
    \includegraphics[width=\textwidth]{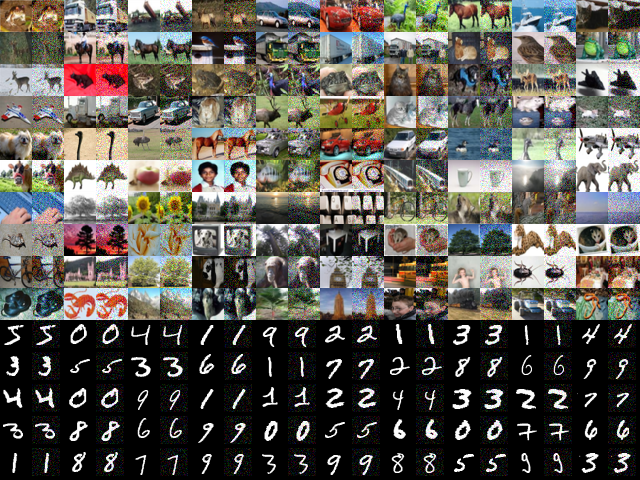}
    \caption{Data reconstructed at the server using DP-Gaussian Noise with \( \sigma^2 \)=0.2 after identification of the malicious layers}
    \label{fig:all_combined_noise}
    \Description{Reconstructed data using DP-Gaussian noise}

\end{figure*}

\begin{figure*}[b!]
    \centering
    \includegraphics[width=\textwidth]{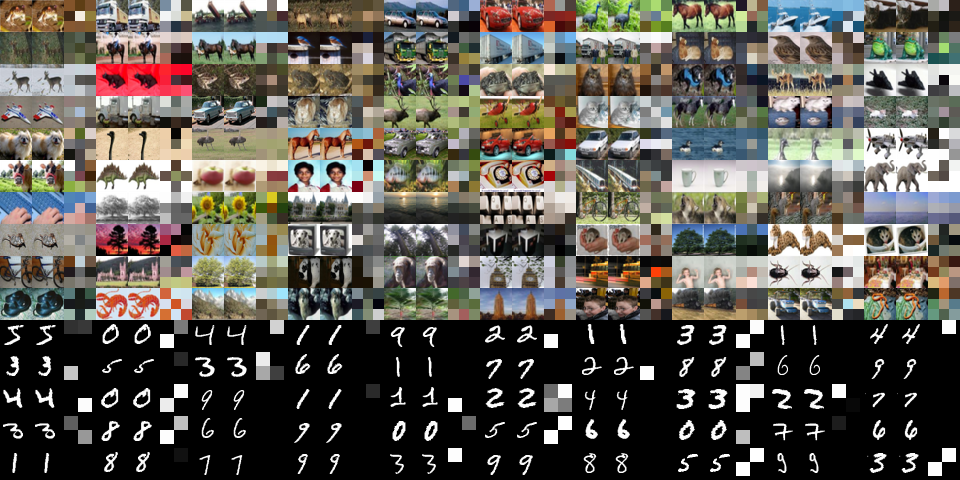}
    \caption{Illustration of reconstruction results across multiple examples. Each group contains three images: the original image (first), the reconstruction produced by the DRA attack using LoKI (second), and the reconstruction generated when pixelated gradients are sent from malicious layers (third) after our detection mechanism.}
    \label{fig:pixelated}
    \Description{Reconstruction results with pixelated gradients}
\end{figure*}

\section{Impact of \DRArmor on Scalability and Computation Overhead}
\label{subsec:scalability}
While evaluating \DRArmor’s practicality in federated settings, it is essential to acknowledge that the computational overhead and scalability of the defense are inherently influenced by multiple factors—including the model’s depth and architecture, the type of dataset used, batch size, and the nature of layers (e.g., convolutional vs. dense). To provide a realistic assessment, we measured per‐iteration wall‐clock time across 200 client nodes using the five datasets described in \S\ref{sec:experiments}, encompassing models of different sizes and complexities. Figure~\ref{fig:runtime-overhead} illustrates the iteration runtimes under four configurations: baseline training (no defense), and DRArmor’s attribution routines (DTD or LRP) with batch sizes of 64 and 128. Notably, these explainability routines are invoked only once per local round rather than per mini-batch, thereby constraining the overhead to a single backward pass (for DTD) or a single relevance propagation (for LRP). \par
Introducing DTD incurs an overhead of approximately 15–25 \% relative to baseline, which can be accommodated within the typical idle windows of federated clients. In contrast, LRP induces a higher latency on deeper architectures—for example, on ImageNet it approaches 1.3 s per iteration. The small fluctuations in each curve capture realistic runtime variability (I/O jitter, network latency, etc.). The expected slowdown when halving the batch size from 128 to 64, on the order of 10–20 \%—is also evident across all methods. \par
Notably, the near‐linear increase in runtime from MNIST’s lightweight model to ImageNet’s deep network demonstrates that \DRArmor’s one‐time, post‐training analysis scales proportionally with model complexity of varying batch sizes. These results confirm that the default DTD‐based detection is efficient and scalable, requiring no specialized hardware, while allowing practitioners to substitute alternative XAI techniques when tighter latency budgets are needed.

\section{Sample reconstructed images for other datasets}

In \S\ref{sec:eval}, we presented a sample reconstruction example demonstrating how \DRArmor mitigates DRA. Here, we extend that analysis to a broader set of examples across the dataset, providing a more comprehensive evaluation of our defense mechanism. 

Figure \ref{fig:all_combined_noise} showcases a reconstruction example where the server attempts to recover input data after applying DP Gaussian noise with \( \sigma^2 \)=0.2, following the identification of malicious layers. This image demonstrates how noise-injected gradients can partially obscure sensitive information while retaining some recognizable structure.

Figure \ref{fig:pixelated} provides a broader set of reconstruction results. Each group of images consists of three components: the original image from the dataset (first), the reconstruction generated by the DRA attack using the LOKI method (second), and the result produced when pixelated gradients from malicious layers are intercepted and modified by our defense mechanism (third). These comparisons highlight the degradation in reconstruction quality achieved through \DRArmor, illustrating its effectiveness in disrupting high-fidelity data recovery by adversaries.

These examples underscore the robustness of \DRArmor in protecting user data in FL settings. By detecting compromised layers and modifying their outputs, our method significantly impairs adversarial reconstructions while maintaining the integrity of the learning process.

\end{document}